\documentclass[10pt,journal,compsoc]{IEEEtran}

\usepackage{cite}
\usepackage{amsthm}
\usepackage{amsmath}
\usepackage{amsfonts}
\usepackage{CJK}
\usepackage{indentfirst}
\usepackage{cases}
\usepackage{url}

\usepackage{graphicx}  
\usepackage{subcaption}
\usepackage{tabularx}
\usepackage{float} 
\usepackage{color}
\usepackage{cases}
\usepackage{ragged2e}
\usepackage[ruled,vlined,linesnumbered,ruled]{algorithm2e}
\usepackage{calligra}
\newtheorem{remark}{Remark}
\usepackage{array}
\usepackage{mdwmath}
\usepackage{mdwtab}
\usepackage{eqparbox}
\usepackage{url}
\usepackage{amsmath}
\usepackage{amssymb}
\usepackage{cite}
\usepackage{amsmath,amsfonts,amssymb}
\usepackage{graphicx}
\usepackage{url}

\begin{document}

\newtheorem{proposition}{Proposition}

\title{Embodied AI-Enhanced Vehicular Networks: An Integrated Large Language Models and Reinforcement Learning Method}

\author{Ruichen Zhang, Changyuan Zhao, Hongyang Du, Dusit Niyato,~\IEEEmembership{Fellow,~IEEE},   Jiacheng Wang, \\
 Suttinee Sawadsitang, 
Xuemin Shen,~\IEEEmembership{Fellow,~IEEE}, and Dong In Kim,~\IEEEmembership{Life Fellow,~IEEE}

\thanks{R. Zhang, C. Zhao, D. Niyato, and J. Wang are with the College of Computing and Data Science, Nanyang Technological University, Singapore (e-mail: ruichen.zhang@ntu.edu.sg, zhao0441@e.ntu.edu.sg, dniyato@ntu.edu.sg, jiacheng.wang@ntu.edu.sg).}

\thanks{H. Du is with the Department of Electrical and Electronic Engineering, University of Hong Kong, Pok Fu Lam, Hong Kong (e-mail: duhy@eee.hku.hk).}

\thanks{S. Sawadsitang is with the College of Arts, Media and Technology, Chiang Mai University, Suthep 50200, Thailand (e-mail: suttinee.s@cmu.ac.th).}



\thanks{X. Shen is with the Department of Electrical and Computer Engineering, University of Waterloo, Canada (e-mail: sshen@uwaterloo.ca).}


\thanks{D. I. Kim is with the Department of Electrical and Computer Engineering, Sungkyunkwan University, Suwon 16419, South Korea (email: dongin@skku.edu).}
}

\markboth{Journal of \LaTeX\ Class Files,~Vol.~14, No.~8, August~2015}%
{Shell \MakeLowercase{\textit{et al.}}: Bare Demo of IEEEtran.cls for Computer Society Journals}

\IEEEtitleabstractindextext{%
\begin{abstract}
\justifying
This paper investigates adaptive transmission strategies in embodied AI-enhanced vehicular networks by integrating large language models (LLMs) for semantic information extraction and deep reinforcement learning (DRL) for decision-making. The proposed framework aims to optimize both data transmission efficiency and decision accuracy by formulating an optimization problem that incorporates the Weber-Fechner law, serving as a metric for balancing bandwidth utilization and quality of experience (QoE). Specifically, we employ the large language and vision assistant (LLAVA) model to extract critical semantic information from raw image data captured by embodied AI agents (i.e., vehicles), reducing transmission data size by approximately more than 90\% while retaining essential content for vehicular communication and decision-making. In the dynamic vehicular environment, we employ a generalized advantage estimation-based proximal policy optimization (GAE-PPO) method to stabilize decision-making under uncertainty. Simulation results show that attention maps from LLAVA highlight the model's focus on relevant image regions, enhancing semantic representation accuracy. Additionally, our proposed transmission strategy improves QoE by up to 36\% compared to DDPG and accelerates convergence by reducing required steps by up to 47\% compared to pure PPO. Further analysis indicates that adapting semantic symbol length provides an effective trade-off between transmission quality and bandwidth, achieving up to a 61.4\% improvement in QoE when scaling from 4 to 8 vehicles.
\end{abstract}

}

\maketitle

\IEEEdisplaynontitleabstractindextext

\IEEEpeerreviewmaketitle

\section{Introduction}

With the advent of the 6G era, the Internet of Vehicles (IoV) is expected to achieve unprecedented advancements, with traffic densities exceeding 0.1--10 Gbps/$\rm m^{2}$ and connection densities reaching up to 10 million devices/$\rm km^{2}$ \cite{9815151}. These improvements will significantly enhance data rates, connectivity, and network capacity, fundamentally transforming IoV services such as real-time navigation, environmental perception, and autonomous decision-making \cite{10571385}. As a key component of IoV, the vehicular network plays a pivotal role in enabling communication between vehicles and infrastructure, facilitating these advanced services. With the growing number of connected vehicles, the demand for these services is increasing, necessitating the deployment of numerous sensors in vehicles and along roadsides to collect and process large amounts of real-time data \cite{wang2024privacy}.

Artificial intelligence (AI) has emerged as a significant technology for enabling vehicles to process this data autonomously and efficiently. Traditional discriminative AI models have been applied to tasks such as object detection \cite{liu2024fedagl} and path planning \cite{wang2023reliable} in vehicular networks. However, as vehicular networks grow more complex, these models struggle to maintain high performance under dynamic conditions and when managing multiple tasks simultaneously \cite{tang2021comprehensive}. As a result, robust AI methods are required to address the increasing complexity of these environments.

Embodied AI has been proposed as a promising solution, which emphasizes the interaction between intelligent systems and their physical surroundings \cite{duan2022survey}. In vehicular networks, embodied AI systems embedded within vehicles enable real-time adaptation to dynamic environments \cite{cunneen2019autonomous}. Equipped with advanced sensors and actuators, these vehicles can perceive their surroundings, make intelligent decisions, and autonomously perform tasks such as obstacle avoidance, traffic management, and collaborative driving. In particular, embodied AI-enhanced vehicular networks rely on two core components, i.e., efficient data processing and adaptive decision-making.
\begin{itemize}
\item For \textbf{data processing}, large language models (LLMs) can be employed to extract critical information via their advanced understanding and generation capabilities \cite{10679152}. Specifically, LLMs analyze multimodal data such as images or videos captured by vehicles, converting them into text-based information that can be more efficiently transmitted across the network. This method leverages multimodal technologies to reduce the amount of raw data that needs to be sent, thereby optimizing network efficiency and bandwidth usage. For example, in \cite{wang2024omnidrive}, an LLM-based data transmission framework was proposed to compress large volumes of sensory data in autonomous driving systems, significantly reducing transmission bandwidth while preserving the quality of information.

\item For \textbf{decision-making}, deep reinforcement learning (DRL) can be utilized to develop adaptive strategies in dynamic environments by continuously adjusting actions based on real-time feedback \cite{10032267}. Specifically, DRL trains agents to make optimal decisions by interacting with their environment and receiving feedback in the form of rewards or penalties. As part of an embodied AI system, this enables vehicles to adjust driving behaviors for tasks such as obstacle avoidance and cooperative driving, ensuring responsiveness and improved performance in complex scenarios \cite{10506539}. For example, in \cite{shao2024semantic}, a DRL-based semantic-aware spectrum-sharing method was proposed to enhance vehicular decision-making, optimizing communication efficiency and safety outcomes in high-speed environments.
\end{itemize}
In addition to their individual strengths, LLMs and DRL complement each other in embodied AI systems. In particular, decision-making often requires continuous interaction between the system and human users. In an embodied AI framework, LLMs act as an optimal interface, enhancing the understanding of contextual and semantic data from multimodal inputs \cite{10531073}. This processed information informs DRL-based decision-making, where the system uses the LLM-generated semantic insights to refine strategies in real-time. By leveraging LLMs, vehicles can convert complex multimodal data into actionable information, which in turn allows DRL to optimize decisions in dynamic vehicular environments.

\begin{figure}[!t]
\centering
\includegraphics[width=0.40\textwidth]{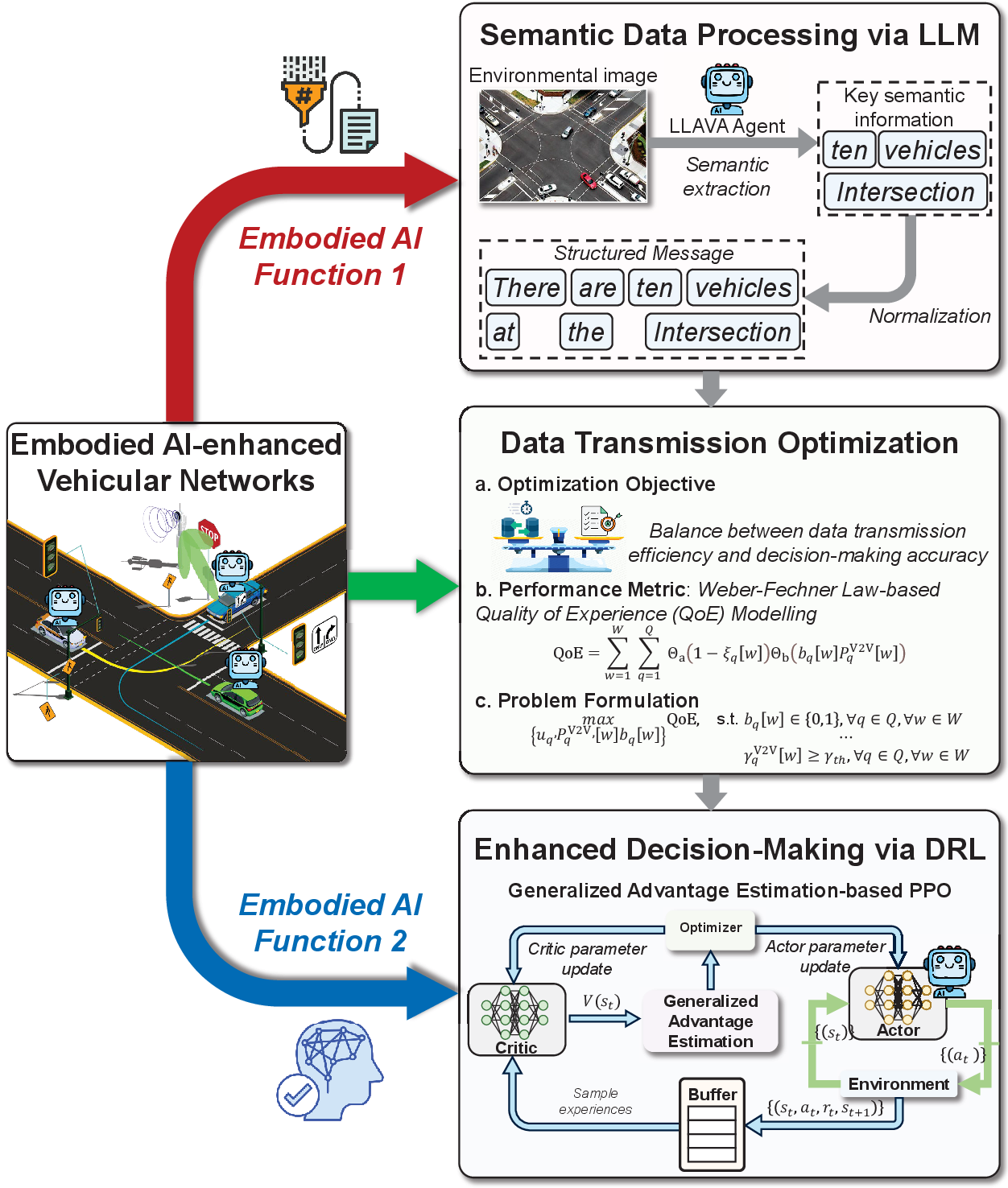}
\caption{The workflow of the proposed embodied AI framework for vehicular networks. The framework comprises two key functions: semantic data processing using LLAVA for efficient information extraction and enhanced decision-making via GAE-PPO to optimize transmission and decision strategies.}
\label{fig:outline}
\end{figure}

Motivated by these insights, we propose a framework that integrates LLMs and DRL within embodied AI-enhanced vehicular networks, as illustrated in Fig.~\ref{fig:outline}. \textit{To the best of the authors' knowledge, this is the first work to explore the joint application of LLMs for data processing and DRL for decision-making in embodied AI-enhanced vehicular networks.} In this work, each vehicle operates as an embodied AI agent, utilizing LLMs to extract and communicate semantic information and employing DRL to optimize decision-making processes. The contributions of this work are summarized as follows: 
\begin{itemize} 
\item \textbf{Data Transmission Formulation}: We formulate an optimization problem aimed at balancing data transmission efficiency and decision-making accuracy. To ensure the quality of transmitted information while managing limited bandwidth, we introduce the Weber-Fechner law as a metric for quantifying quality of experience (QoE). This law provides an objective function to ensure that the perceived quality of transmitted data remains high, even as bandwidth constraints increase.

\item \textbf{Semantic Data Processing via LLM}: We leverage the large language and vision assistant (LLAVA) to extract semantic information from raw image data. Specifically, LLAVA converts visual inputs into structured text representations using its cross-modal alignment capability, which integrates a visual encoder, a projection matrix, and a BERT-based language model. This process significantly reduces the transmission bandwidth while preserving essential contextual details needed for vehicular communication and decision-making.

\item \textbf{Enhanced Decision-Making via DRL}: We propose a generalized advantage estimation-based proximal policy optimization (GAE-PPO) to improve decision-making in dynamic vehicular environments. The GAE-PPO method incorporates generalized advantage estimation to reduce variance in policy gradient updates, stabilizing the training process. It also employs policy clipping and replay buffers to enhance convergence and robustness, ensuring that the decisions made by our embodied AI agents align with QoE optimization goals. \end{itemize}

Simulation results show that the proposed embodied AI framework significantly outperforms the baselines in multiple aspects. The attention maps generated by LLAVA indicate that the model accurately focuses on relevant image regions during semantic extraction, effectively identifying key features such as cars and parking spaces. Compared to other vision-LLM models (e.g., LLAVA-1.5-13b-hf, Qwen-VL-Chat, Deepseek-vl-7b-base, and Moondream2), LLAVA-1.5-7b-hf achieves superior accuracy while maintaining lower inference time, outperforming larger models like LLAVA-1.5-13b-hf by reducing inference time by 40\% on average and demonstrating a smaller parameter size, making it computationally efficient. The analysis of achievable QoE as a function of the number of embodied AI agents demonstrates that our method maintains superior QoE under increasing network load compared to conventional methods, with a 36\% improvement over DDPG and a 25.2\% gain when scaling from 12 to 16 vehicles. Additionally, the convergence analysis reveals that the GAE-PPO method ensures faster and more stable convergence, reducing convergence steps by up to 54 compared to pure PPO. Finally, an investigation into the relationship between semantic symbol length and transmission quality confirms that adapting the symbol length provides an effective trade-off between communication quality and bandwidth utilization, achieving up to a 61.4\% improvement in QoE when scaling from 4 to 8 vehicles.

The rest of this paper is organized as follows: Section II reviews the related work. Section III presents the system model, including the details of LLAVA-based semantic extraction and GAE-PPO for optimization. Section IV discusses the proposed framework for vehicular network communication. Section V provides the simulation results and performance analysis, demonstrating the advantages of our method over conventional methods. Finally, Section VI concludes the work.

\section{Related Work}
In this section, we review the relevant literature across two key domains, i.e., vehicular Networks and embodied AI. We highlight significant advancements and identify the gaps that our work aims to address.

\subsection{Vehicular Networks}
Recent advancements in vehicular networks have garnered significant attention due to their potential to improve road safety, optimize resource allocation, and enhance communication reliability. For instance, Liang et al. \cite{liang2019spectrum} proposed a multi-agent reinforcement learning (MARL) framework for spectrum sharing between V2V and V2I communications, where their method improved spectrum efficiency by allowing vehicles to manage spectrum resources in high-mobility environments cooperatively. Additionally, Liang et al. \cite{liang2018graph} proposed a resource-sharing strategy leveraging graph-based methods to optimize vehicular network performance in dynamic environments, where performance was measured through reduced interference and enhanced communication reliability. Xue et al. \cite{xue2024cooperative} introduced DRL to tackle power allocation in vehicular networks, focusing on improving ultra-reliable low-latency communication (URLLC) while reducing packet duplication, which directly impacted communication reliability and overall system latency. Similarly, Zhang et al. \cite{10643168} developed a federated learning-based method for gradient quantization in vehicle edge computing networks. This method improved communication efficiency by minimizing bandwidth consumption during model updates and enhanced resource sharing across distributed vehicular nodes. Furthermore, Yao et al. \cite{yao2023secure} proposed a DRL-based secure transmission scheme for V2V communications using joint radar and communication systems. Their method protected V2V links from eavesdropping and interference while maintaining communication integrity, spectral efficiency, and signal-to-noise ratio (SNR) even in high-mobility environments.

Despite these significant contributions, most existing research primarily focused on traditional performance metrics like spectrum efficiency, latency, and security. However, new metrics such as semantic data transmission and decision-making efficiency become essential when considering embodied AI. Additionally, when incorporating LLMs for data extraction and processing, the QoE must be considered. Our work addresses this by introducing the Weber-Fechner law as a novel metric for balancing bandwidth and QoE in data transmission for vehicular networks.

\subsection{Embodied AI}
With the rapid evolution of AI, the concept of embodied AI has emerged, emphasizing the interaction between physical entities and intelligent systems. The term ``embodied AI" refers to AI systems embedded within physical objects, enabling them to perceive, interpret, and respond to their environment\cite{savva2019habitat}.  The related work on embodied AI can be broadly divided into two primary areas, i.e., data extraction and decision-making.

For data processing, most of the studies have focused on utilizing LLMs to efficiently process and transmit critical information, reducing the need for raw data transmission. For instance, Song et al. \cite{song2023llm} proposed an embodied agents framework that used an LLM-based method for few-shot grounded planning in embodied agents. Their method allowed the system to extract key information from multimodal inputs, enabling agents to dynamically update plans based on minimal data, significantly reducing the volume of sensory data required for transmission.  Similarly, in \cite{zhang2024towards}, Zhang et al. introduced a multi-embodied system that enhances LLMs' ability to extract relevant data for multi-agent collaboration. Their method improved task execution by refining the plan based on environmental inputs. This led to more efficient data extraction and reduced the amount of data exchanged between agents and their environment. Moreover, Mower et al.\cite{mower2024ros} developed a framework that integrates LLMs with embodied robotic systems, where data extraction is optimized through task feedback and structured reasoning. By leveraging LLMs to extract essential task information from multimodal data, the framework reduced the need to transmit large volumes of raw sensory input, enhancing the efficiency of robotic decision-making and communication.  Additionally, in \cite{zhang2024badrobot}, Zhang et al. explored the vulnerabilities of multimodal LLMs in embodied AI, focusing on extracting relevant actions from complex, multimodal data and demonstrating how these systems can reduce unnecessary transmissions by filtering unsafe actions, underscoring the importance of secure data extraction for efficient operation in embodied AI systems.

\begin{figure*}[t]
\centering
\includegraphics[width=0.95\textwidth]{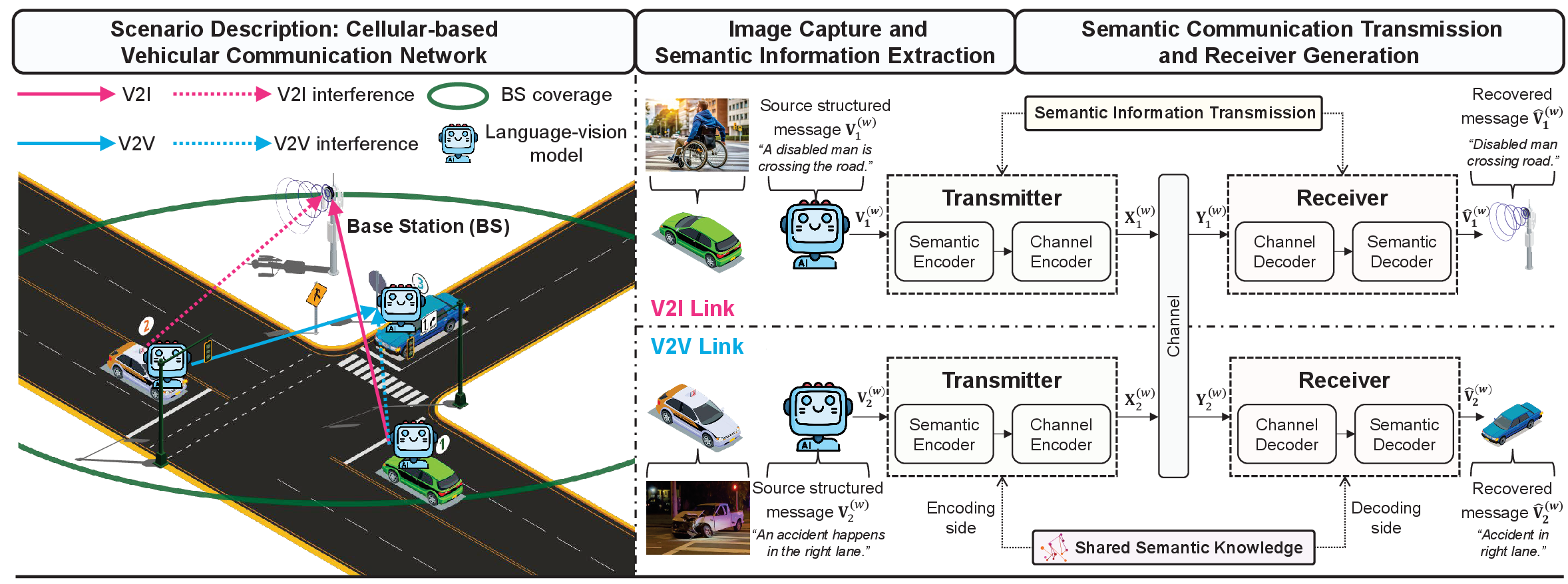}
\caption{{System model illustrates a cellular-based vehicular communication network, where embodied AI vehicles utilize semantic communication to encode and decode structured messages for efficient and reliable data exchange \cite{Shunpu_SemCom}.}}
\label{fig:enter-label}
\end{figure*}

For decision-making, most of the research in embodied AI has focused on using DRL to develop strategies in dynamic environments.  For instance, Ying et al. \cite{ying2024peac} introduced the PEAC algorithm, which used RL to pre-train embodied agents across multiple embodiments. By learning task-agnostic strategies, PEAC improved decision-making efficiency and adaptability in handling dynamic environments. Similarly, in \cite{10146515}, Long et, al. developed a human-in-the-loop simulation embodied platform that integrated RL to learn control policies from human demonstrations, enhancing decision-making efficiency in surgical tasks. Moreover, Liu et al. \cite{liu2023simple} explored language learning as a byproduct of meta-reinforcement learning, showing how agents indirectly learned language cues through decision-making tasks, further highlighting RL’s role in extracting and utilizing environmental data for task performance improvement. Additionally, in \cite{tan2024true}, Tan et al. proposed the TWOSOME framework, combining LLMs and DRL for decision-making in embodied environments. Using PPO to interact with tasks, they optimized decision-making and reduced the need for predefined datasets by enabling real-time data extraction and optimization.

Despite these significant contributions, most existing works on embodied AI primarily focus on improving algorithmic efficiency and decision-making without fully exploring its potential for resource optimization in vehicular networks. Additionally, while LLMs and DRL have been successfully applied in various embodied AI systems, their integration in optimizing both communication and decision-making processes in vehicular networks remains underexplored. Therefore, this work aims to fill this gap by leveraging the strengths of embodied AI, LLMs, and DRL to enhance both resource optimization and decision-making within the context of vehicular networks.

\section{System Model}\label{system}
As illustrated in Fig.~\ref{fig:enter-label}, we consider a cellular-based vehicular communication network in an urban environment, where $I$ vehicles equipped with embodied AI systems drive within the communication range of a base station (BS). {To associate a vehicle with a BS, the system selects the link based on the strongest received signal strength indicator (RSSI) value at the BS \cite{10041763}.} The network comprises $W$ vehicle-to-infrastructure (V2I) links and $Q$ vehicle-to-vehicle (V2V) links, which support both direct communication between vehicles and communication with the BS. {The V2I links are used for transmitting high-priority or aggregated data to the BS, while V2V links facilitate direct communication between nearby vehicles, reducing network load and latency in localized areas.} For clarity, we use $i \in \{1, 2, \ldots, I\}$ to denote the index of vehicle, $w \in \{1, 2, \ldots, W\}$ to denote the index of V2I link, and $q \in \{1, 2, \ldots, Q\}$ to denote the index of V2V link, respectively.

\subsection{Image Capture and Semantic Information Extraction}
In this model, each vehicle is equipped with onboard sensors that capture real-time environmental images, denoted by $\mathbf{I}_i$. These images are processed by a vision-LLM model, i.e., LLAVA, to extract relevant semantic information and convert it into textual descriptions. {Each vehicle uses its onboard embodied AI system to process and extract semantic information locally, performing that computational tasks are offloaded from the BS and minimizing the volume of raw data that needs to be transmitted.} The extracted semantic information $\mathbf{M}_i$ contains details about other vehicles, pedestrians, road conditions, and traffic signals. 

The semantic processing system within each vehicle utilizes the textual information $\mathbf{M}_i$ to generate a structured message, denoted by $\mathbf{V}_i^{(w)}$, where $w$ indicates the specific V2I link over which the information will be transmitted. The structured message $\mathbf{V}_i^{(w)}$ consists of $l_{\rm max}$ words, i.e., 
\begin{equation}\label{eq2}
\mathbf{V}_i^{(w)} = \left[ v_{i,1}^{(w)}, v_{i,2}^{(w)}, v_{i,l}^{(w)}, \ldots, v_{i,l_{\rm max}}^{(w)} \right],
\end{equation}
{where $v_{i,l}^{(w)}$ represents the $l$-th word in the structured message for the $i$-th vehicle on link $w$. This message is then passed through the semantic encoder of the vehicle's transmitter, which extracts and encodes the underlying semantic information into a symbol vector $\mathbf{S}_i^{(w)}$. The encoding process can be formalized as
\begin{equation}\label{eq3_corrected} 
\mathbf{S}_i^{(w)} = \text{SE}_{\alpha}\left( \mathbf{V}_i^{(w)} \right), 
\end{equation}
{where $\text{SE}_{\alpha}(\cdot)$ represents the semantic encoding function with a set of hyper-parameters $\alpha$.} The resulting semantic symbol vector $\mathbf{S}_i^{(w)}$ is then represented as 
\begin{equation}
\mathbf{S}_i^{(w)} = \left[ s_{i,1}^{(w)}, s_{i,2}^{(w)}, s_{i,u}^{(w)}, \ldots, s_{i,u_{\rm max}}^{(w)} \right], 
\end{equation}
{where $u$ denotes the average number of semantic symbols used per word with the unit of \textit{semantic unit}.}

\subsection{SemCom Transmission and Reconstruction}
After the semantic information is encoded into a symbol vector $\mathbf{S}_i^{(w)}$, it is transmitted over the wireless channel. The transmitted signal for each V2I or V2V link is denoted as:
\begin{equation}\label{eq5_corrected} 
\mathbf{X}_i^{(w)} = \text{CE}_{\beta}\left( \mathbf{S}_i^{(w)} \right), 
\end{equation}
where $\text{CE}_{\beta}(\cdot)$ denotes the channel encoding function parameterized by $\beta$. The expression in (\ref{eq5_corrected}) represents the process of preparing the semantic symbols for transmission over the corresponding $w$-th link.

Each V2I link is allocated a specific frequency subband, while the $Q$ V2V links may share these subbands, leading to potential interference. The signal-to-interference-plus-noise ratios (SINRs) for the $w$-th V2I link and the $q$-th V2V link on the $w$-th subband are respectively denoted as
\begin{equation}\label{eq7_corrected} 
\gamma_w^{\text{V2I}} = \frac{P_w^{\text{V2I}} h_{w,\text{BS}}^{(w)}}{\sigma^2_{\rm a} + \sum_{q=1}^{Q} b_q[w] P_q^{\text{V2V}}[w] h_{q,\text{BS}}^{(w)}}, 
\end{equation}
and
\begin{equation}\label{eq8_corrected} 
\gamma_q^{\text{V2V}}[w] = \frac{P_q^{\text{V2V}}[w] h_{q,q}^{(w)}}{\sigma^2_{\rm b} + P_w^{\text{V2I}} h_{w,q}^{(w)} + \sum_{\substack{q' \ne q}}^{Q} b_{q'}[w] P_{q'}^{\text{V2V}}[w] h_{q',q}^{(w)}}, 
\end{equation}
where $P_w^{\text{V2I}}$ and $P_q^{\text{V2V}}[w]$ represent the transmission powers for the $w$-th V2I and $q$-th V2V links, respectively. $\sigma_{\rm a}^2$ and $\sigma_{\rm b}^2$ denote the corresponding Gaussian noise, respectively. The binary variable $b_q[w] = 1$ indicates that the $q$-th V2V link shares the $w$-th V2I link\footnote{{The interaction between V2V and V2I links is crucial for supporting dynamic data exchange in vehicular networks. Specifically, the V2V links facilitate direct, localized communication between vehicles, reducing latency. In contrast, V2I links are used for relaying high-priority or aggregated data from vehicles to the BS, providing centralized control and data management. The V2V and V2I communications are distinct. That is, V2V transmissions are not forwarded to the BS via V2I links. Such a system setting has been adopted in existing works, e.g., \cite{huang2024joint} and \cite{9808399}.}}. {The channel gains $h_{w,\text{BS}}^{(w)}$, $h_{q,\text{BS}}^{(w)}$, $h_{q,q}^{(w)}$, $h_{w,q}^{(w)}$, and $h_{q',q}^{(w)}$ represent both small-scale fading and large-scale path loss effects}, affecting the quality of the signal for both V2I and V2V links. The received semantic signal at the receiver, managed by the embodied AI agents, is expressed as
\begin{equation}
{\mathbf{Y}_i^{(w)} = \textbf{H}_i^{(w)} \mathbf{X}_i^{(w)} + \mathbf{N},}
\end{equation}
where $\textbf{H}_i^{(w)}$ represents the overall channel gain matrix for the $w$-th V2I and V2V links, and $\mathbf{N}$ denotes the noise vector.

Upon receiving the transmitted semantic symbols, the vehicle receiver, controlled by the embodied AI system, decodes and reconstructs the original semantic information using channel and semantic decoding techniques. The decoded signal is represented as
\begin{equation}\label{eq10_v1}
\hat{\mathbf{S}}_i^{(w)} = \text{SE}_{\mu}^{-1}\left( \text{Enc}_{\nu}^{-1}\left( \mathbf{Y}_i^{(w)} \right) \right),
\end{equation}
where $\hat{\mathbf{S}}_i^{(w)}$ denotes the recovered semantic symbol vector for the $i$-th vehicle on the $w$-th link, $\text{SE}_{\mu}^{-1}(\cdot)$ is the semantic decoder with parameter $\mu$, and $\text{Enc}_{\nu}^{-1}(\cdot)$ is the channel decoder with parameter $\nu$. {The $\mu$ and $\nu$ consist of network settings of the semantic decoder and channel decoder, respectively.}

To assess the accuracy of the reconstructed information, the cross-entropy (CE) loss function is adopted \cite{xie2021deep}, as shown in (\ref{chargedpower}), where $q\left(v_{i,l}^{(w)}\right)$ is the actual probability of the $l$-th word in the original message $\mathbf{V}_i^{(w)}$, and $p\left(v_{i,l}^{(w)}\right)$ is the predicted probability in the recovered message $\hat{\mathbf{V}}_i^{(w)}$.

Additionally, mutual information (MI) between the transmitted and received symbols is evaluated as in (\ref{eq12}), where $p\left(\mathbf{X}_i^{(w)}\right)$ and $p\left(\mathbf{Y}_i^{(w)}\right)$ are the marginal probabilities of the transmitted and received symbols, respectively. This metric provides insight into the efficiency of semantic communication, helping to evaluate how effectively the transmitted information is preserved through the communication channel.

\begin{figure*}
\begin{equation}
\begin{aligned}
\mathcal{L}_{CE}\left(\mathbf{V}_i^{(w)}, \hat{\mathbf{V}}_i^{(w)}; \alpha, \beta, \mu, \nu\right)   = -\sum_{l=1}^{l_{\rm max}} q\left(v_{i,l}^{(w)}\right) \log p\left(v_{i,l}^{(w)}\right) + \left(1 - q\left(v_{i,l}^{(w)}\right)\right) \log\left(1 - p\left(v_{i,l}^{(w)}\right)\right).
\label{chargedpower}
\end{aligned}
\end{equation}
\end{figure*}

\begin{figure*}
\begin{equation}
\label{eq12}
\begin{aligned}
I\left(\mathbf{X}_i^{(w)}; \mathbf{Y}_i^{(w)}\right)  = \sum_{\mathbf{X}_i^{(w)}, \mathbf{Y}_i^{(w)}} & p\left(\mathbf{X}_i^{(w)}, \mathbf{Y}_i^{(w)}\right) \log \frac{p\left(\mathbf{X}_i^{(w)}, \mathbf{Y}_i^{(w)}\right)}{p\left(\mathbf{X}_i^{(w)}\right) p\left(\mathbf{Y}_i^{(w)}\right)}
=\mathbb{E}{_{p(x,y)}\Big[\log\frac{p\left(\mathbf{X}_i^{(w)}, \mathbf{Y}_i^{(w)}\right)}{p\left(\mathbf{X}_i^{(w)}\right) p\left(\mathbf{Y}_i^{(w)}\right)}\Big]}.
\end{aligned}
\end{equation}
\hrule
\end{figure*}

Using the dual representation of the Kullback–Leibler (KL) divergence \cite{10753492}, we can derive a lower bound for the MI as follows:
\begin{equation}
I\left(\mathbf{X}_i^{(w)}; \mathbf{Y}_i^{(w)}\right) \geq \mathcal{L}_{MI}\left(\mathbf{X}_i^{(w)}, \mathbf{Y}_i^{(w)}; \alpha, \beta, T\right),
\end{equation}
where $\mathcal{L}_{MI}$ {is the loss function used to train the neural network to get $\alpha$, $\beta$ and $T$ for the MI}, and $T$ represents the temperature parameter that  $T$ controls the sharpness of the probability distribution, effectively adjusting the confidence level of the model's predictions during training.

\subsection{Performance Metrics and Formulated Problem}
{To evaluate the effectiveness of the reconstructed semantic information transmitted through V2I and V2V links}, we compute the cosine similarity between the bidirectional encoder representation of Transformers (BERT) embeddings of the original and reconstructed messages \cite{10746594}. {The cosine similarity describes the sentence semantic similarity between the recovered $\hat{\mathbf{V}}_i^{(w)}$ and actual $\mathbf{V}_i^{(w)}$, which reflects whether the semantic of the sentences are correctly t transmitted.} Specifically, the cosine similarity is defined as
\begin{equation}\label{eq15}
\xi  = \frac{\left|B\left( \mathbf{V}_i^{(w)} \right) \cdot B\left( \hat{\mathbf{V}}_i^{(w)} \right)\right|}{\left\| B\left( \mathbf{V}_i^{(w)} \right) \right\| \left\| B\left( \hat{\mathbf{V}}_i^{(w)} \right) \right\|},
\end{equation}
where $B\left(\cdot\right)$ represents the BERT model used to generate feature vectors for the sentences. The value $\xi$ ranges from 0 to 1, with $\xi = 1$ indicating perfect similarity and $\xi = 0$ indicating no similarity.

To optimize the performance of our vehicular communication system, we introduce a QoE metric inspired by the Weber-Fechner law, which models the relationship between the magnitude of a stimulus and its perceived intensity  \cite{10654734}. The QoE is influenced by the semantic similarity (i.e., $\xi_q[w]$) between transmitted and received images, which depends on both the number of semantic symbols and the SINR of the communication link. The QoE can be expressed as
\begin{equation}
\text{QoE} = \sum_{w=1}^W \sum_{q=1}^Q \Theta_{\rm a}\left(1 - \xi_q[w]\right)  \Theta_{\rm b}\left( b_q[w] P_q^{\text{V2V}}[w] \right), 
\end{equation}
where $\Theta_{\rm a}(\cdot)$ and $\Theta_{\rm b}(\cdot)$ are functions that normalize the impact of semantic similarity $\xi_q[w]$ and the transmission power $P_q^{\text{V2V}}[w]$ on the overall QoE. {To ensure clarity, the QoE metric integrates both the semantic accuracy of reconstructed messages and the efficiency of resource utilization. This dual consideration helps strike a balance between the quality of the transmitted information and the optimal use of network resources.}

{The optimization problem focuses on achieving high QoE while ensuring efficient resource utilization.} In this context, the embodied AI agents (i.e., vehicles) play a crucial role in generating and adapting communication strategies. These agents dynamically optimize these parameters, i.e., transmission power, semantic symbol allocation, and channel usage, based on the real-time analysis of the environment and communication needs. Thus, the optimization problem is formulated as follows:
\begin{subequations}\label{P1_revised}
    \begin{flalign}
    \max_{\{P_q^{\rm V2V}[w], b_q[w], \ u_q\}} & {\rm QoE}, \label{P1_revised_a1} \\
    {\rm s.t.} \quad & b_q[w] \in \{0, 1\}, \quad \forall q \in Q, \forall w \in W, \label{P1_revised_b1} \\
    & \sum_{w=1}^W b_q[w] \leq 1, \quad \forall q \in Q, \label{P1_revised_c1_1} \\
    & \sum_{q=1}^Q b_q[w] \leq 1, \quad \forall w \in W, \label{P1_revised_c1_2} \\
    & u_q \in \{1, \ldots, u_{q_{\text{max}}}\}, \forall q \in Q, \label{P1_revised_d1} \\
    & \xi_q[w] \geq \xi_{\text{th}}, \quad \forall q \in Q, \forall w \in W, \label{P1_revised_e1} \\
    & P_q^{\rm V2V}[w] \in [P_{\rm min}, P_{\rm max}], \forall q \in Q, \forall w \in W, \label{P1_revised_f1} \\
    & \gamma_q^{\rm V2V}[w] \geq \gamma_{\rm th}, \quad \forall q \in Q, \forall w \in W. \label{P1_revised_g1}
    \end{flalign}
\end{subequations}
Here,  constraint (\ref{P1_revised_b1}) ensures that each V2V link either shares or does not share a V2I link. Constraints (\ref{P1_revised_c1_1}) and (\ref{P1_revised_c1_2}) ensure that each vehicle and each subband are allocated appropriately, i.e., no overlapping resource usage occurs.  {Constraint (\ref{P1_revised_d1}) specifies the range of the semantic symbol allocation $u_q$.} Constraint (\ref{P1_revised_e1}) ensures that the semantic similarity $\xi_q[w]$ remains higher than or equal to the minimum threshold $\xi_{\text{th}}$ to maintain communication quality. Constraints (\ref{P1_revised_f1}) and (\ref{P1_revised_g1}) limit the transmission power and require the SINR to be higher than or equal to the threshold $\gamma_{\text{th}}$ to ensure efficient communication.

{It is observed that Problem (\ref{P1_revised}) is non-convex and involves combinatorial aspects due to constraints such as (\ref{P1_revised_b1}), (\ref{P1_revised_c1_1}), and (\ref{P1_revised_c1_2}). This complexity, combined with the highly dynamic nature of vehicular networks, presents a significant challenge. Specifically, traditional optimization algorithms often fall short in adapting to rapid changes in channel state or network state, requiring repeated re-execution of computationally expensive procedures. } {To address these issues, we propose an embodied AI framework that integrates LLAVA and PPO to solve it in the next section.}

\section{Proposed Embodied AI Framework}

In this section, we present an embodied AI framework in which LLAVA is utilized for semantic information extraction from environmental images, while PPO is employed to dynamically optimize communication parameters in response to environmental changes.

\subsection{LLAVA-Based Semantic Information Extraction}

To enable the extraction of semantic information from images captured by the embodied AI agents (i.e., vehicles), we employ the LLAVA model. First, we provide an overview of LLAVA, followed by details on its design for semantic data generation.

\subsubsection{Overview of LLAVA}

LLAVA is a large multimodal model designed to act as a general-purpose visual assistant by incorporating the power of LLMs and visual encoders such as contrastive language–image pre-training \cite{xie2023ra}. The model is trained to align image features with language instructions, enabling multimodal tasks such as image understanding, visual reasoning, and conversational interaction with visual content. The LLAVA architecture uses linear projections to connect visual and language modalities, making it a simple yet effective framework for visual-language alignment \cite{liu2024visual}. Specifically, given images $\mathbf{I}_i$ captured by the $i$-th vehicle through its onboard sensors, LLAVA uses a visual encoder denoted as $g(\cdot)$ to extract visual features $\mathbf{Z}_i = g(\mathbf{I}_i)$. These visual features are then projected into the language model’s word embedding space using a linear transformation. The architecture of LLAVA is shown in Fig.~\ref{fig:LLAVA}, where this process effectively maps visual information into a language-interpretable format, allowing LLAVA to respond to visual queries with natural language output. 
\begin{figure}[!t]
\centering
\includegraphics[width=0.49\textwidth]{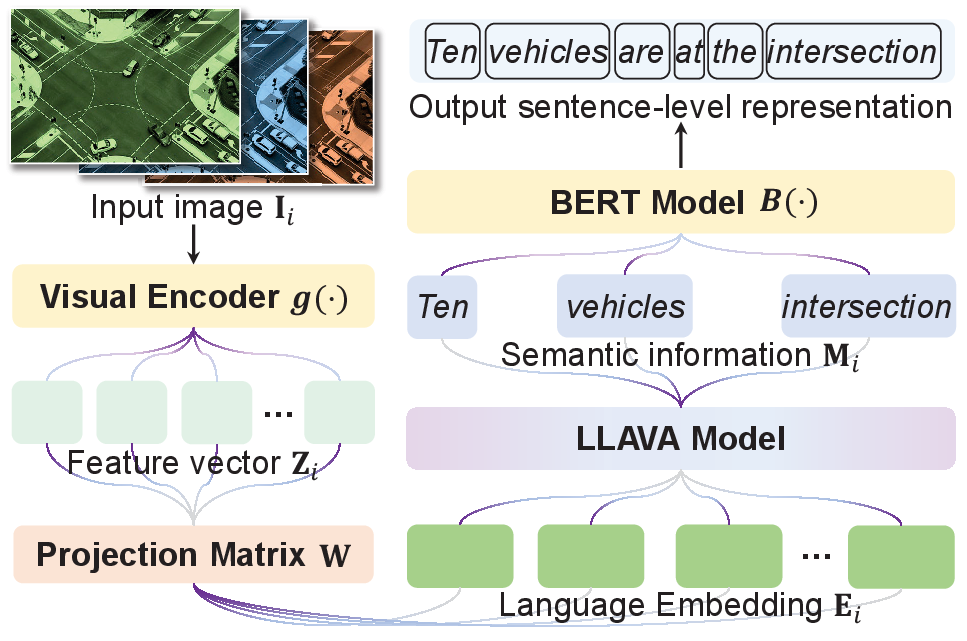}
\caption{The architecture of the LLAVA model for semantic extraction and language embedding. The input image $I_i$ is processed by the visual encoder $g(\cdot)$ to generate feature vectors, which are then transformed through a projection matrix $W$ and processed by the LLAVA model to extract semantic information $M_i$. The BERT model $B(\cdot)$ is used to generate the final sentence-level representation, capturing key information from the visual input.}
\label{fig:LLAVA}
\end{figure}

\subsubsection{LLAVA Design for Extraction}

The proposed LLAVA design in this work aims to support real-time vehicular communication scenarios. The model processes captured images to generate semantic information used for subsequent decision-making. The LLAVA model is employed in the following steps:

\textbf{Data Generation Using LLAVA:}  
Given input images $\mathbf{I}_i$, LLAVA extracts semantic information $\mathbf{M}_i$. Initially, the contrastive language–image pre-training visual encoder is applied to convert the image into feature vectors, i.e.,
\begin{equation}
    \label{z_i}
    \mathbf{Z}_i = g(\mathbf{I}_i),
\end{equation}
where $g(\cdot)$ denotes the contrastive language–image pre-training visual encoder, and $\mathbf{Z}_i$ represents the visual features extracted from the vehicle’s environment.

Next, these features are projected into the word embedding space of the language model by applying a trainable linear projection matrix $\mathbf{W}$, i.e.,
\begin{equation}
 \label{h_i}
    \mathbf{E}_i = \mathbf{W} \cdot \mathbf{Z}_i,
\end{equation}
where $\mathbf{E}_i$ represents the language embedding for the image features. The semantic information $\mathbf{M}_i$ for the embodied AI agent is then derived as:
\begin{equation}
 \label{m_i}
    \mathbf{M}_i = \text{LLAVA}(\mathbf{I}_i; \theta_i),
\end{equation}
where $\theta_i$ represents the parameters of the LLAVA model.

\textbf{LLAVA Model Fine-Tuning:}  
To adopt LLAVA to our vehicular network scenario, the model is fine-tuned using training data generated from real-world driving scenarios. The goal of fine-tuning is to minimize the semantic mismatch between the ground-truth semantic information $\mathbf{M}_i$ and the predicted semantic output $\hat{\mathbf{M}}_i$. The loss function $\mathcal{L}$ used for fine-tuning can be expressed as
\begin{equation}
 \label{l_i}
    \min_{\theta} \sum_{i=1}^{I} \mathcal{L}\left(\mathbf{M}_i, \hat{\mathbf{M}}_i\right) = \min_{\theta} \sum_{i=1}^{I} \|\mathbf{M}_i - \hat{\mathbf{M}}_i\|^2,
\end{equation}
where $\theta$ denotes the trainable parameters of LLAVA.

\begin{remark}
This loss function ensures that the generated semantic information $\mathbf{M}_i$  aligns as closely as possible with the actual environment context perceived by the embodied AI agents, improving the accuracy of transmitted messages.
\end{remark}

The connection between the visual features and word embeddings is further refined through a CE loss during training, i.e.,
\begin{equation}
 \label{l_ce}
    \mathcal{L}_{CE}\left(\mathbf{M}_i, \hat{\mathbf{M}}_i\right) = \sum_{l=1}^{l_{\rm max}} q\left(v_{i,l}\right) \log p\left(v_{i,l}\right),
\end{equation}
where $q(v_{i,l})$ is the true probability of the word $v_{i,l}$, and $p(v_{i,l})$ is the predicted probability in the extracted semantic message $\mathbf{M}_i$.

\textbf{Semantic Information Representation:}  
The semantic information $\mathbf{M}_i$ generated by LLAVA is encoded using the BERT model to obtain a sentence-level representation, denoted as $B(\mathbf{M}_i)$. {The quality of the transmitted semantic content is evaluated using (\ref{eq15}) to ensure the accuracy of the transmitted and received information.}

\begin{figure*}[!t]
\centering
\includegraphics[width=0.95\textwidth]{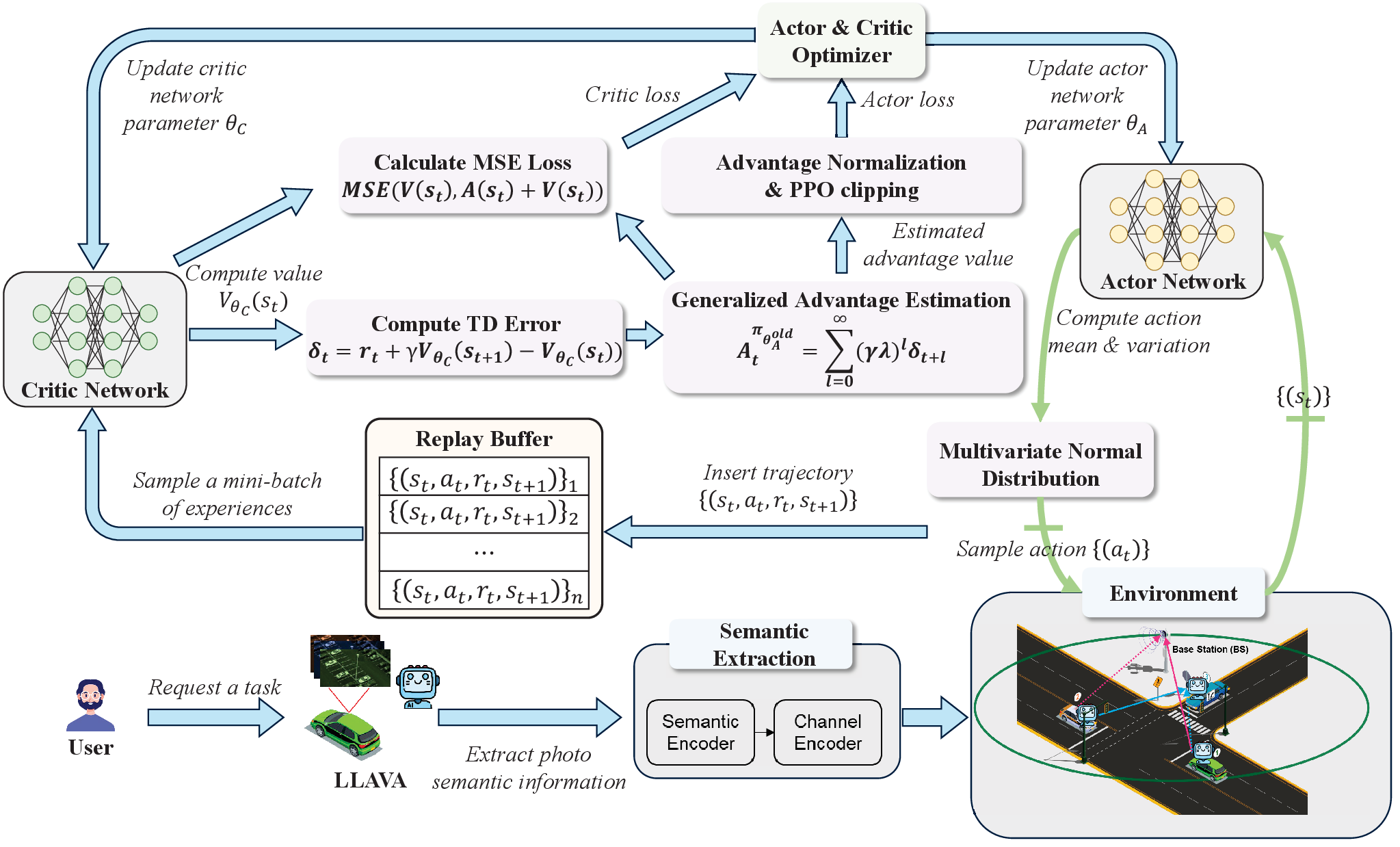}
\caption{The workflow of the GAE-PPO method for optimizing transmission strategies. The actor and critic networks are updated iteratively through a combination of replay buffer sampling, temporal-difference (TD) error calculation, advantage estimation, and policy clipping. The actor network generates actions using a multivariate normal distribution, while the critic network evaluates state values, contributing to the stability and convergence of the learning process.}
\label{fig:DRL}
\end{figure*}

{In such a manner, key information captured by vehicles can be processed as semantic information through LLAVA.} The overall algorithm for semantic information extraction and representation using LLAVA is summarized in Algorithm \ref{alg:llava_extraction}. 

\begin{algorithm}[h]
    \caption{{{LLAVA-Based Semantic Information Extraction for Vehicular Networks.}}} \label{alg:llava_extraction}
    $\mathbf{Input:}$ Input image $\mathbf{I}_i$ from vehicle $i$;  contrastive language–image pre-training encoder $g(\cdot)$; LLAVA model parameters $\theta$; projection matrix $\mathbf{W}$; BERT model for representation;\\
    
    Use pretrained contrastive language–image pre-training encoder to obtain visual features from the input image via (\ref{z_i}); \quad \text{\textit{\# Extract Visual Features}} \\

    Project the extracted visual features into the word embedding space using a trainable linear projection via (\ref{h_i}); \quad \text{\textit{\# Project Features to Embedding Space}} \\

    Utilize LLAVA to generate semantic information from the projected features via (\ref{m_i});  \quad \text{\textit{\# Derive Semantic Information Using LLAVA}} \\

    Fine-tune LLAVA with real-world vehicular data to minimize semantic mismatch via (\ref{l_i}); \quad \text{\textit{\# Fine-Tune LLAVA for Scenario Adaptation}} \\

    Encode the generated semantic information using BERT to obtain a sentence-level representation  $B(\mathbf{M}_i)$;  \quad \text{\textit{\# Semantic Representation with BERT}} \\

    $\mathbf{Output:}$ Extracted semantic information $\mathbf{M}_i$\;
\end{algorithm}

\subsection{PPO-Based Transmission Strategy}

{The proposed PPO-based transmission strategy begins with a service request processed by LLAVA. LLAVA interprets the request and extracts relevant semantic information from images captured by vehicle-mounted cameras.} After extracting the semantic information via LLAVA, an efficient transmission strategy is required for vehicular networks. To address this, we propose a PPO-based method to optimize the transmission policy.

\subsubsection{Overview of PPO}

PPO is a policy gradient method within the actor-critic framework, where the actor network defines the transmission policy (parameterized by $\theta_A$), and the critic network estimates the value function (parameterized by $\theta_C$). {PPO aims to optimize the policy to maximize expected rewards while ensuring smooth policy transitions to prevent drastic and destabilizing updates \cite{ji2022trajectory}. The core of PPO lies in its use of a surrogate objective function that encourages updates to the policy without large shifts from the previous policy $\pi_{\theta_A^{\text{old}}}$.} The surrogate objective function at each time step $t$ is defined as
\begin{equation}
J(\theta_A) = \mathbb{E}_t \left[ \rho_t(\theta_A) A_t^{\pi_{\theta_A^{\text{old}}}} \right],
\end{equation}
where $\rho_t(\theta_A) = \frac{\pi_{\theta_A}(a_t|s_t)}{\pi_{\theta_A^{\text{old}}}(a_t|s_t)}$ represents the probability ratio between the new and old policies, and $A_t^{\pi_{\theta_A^{\text{old}}}}$ is the advantage function at time $t$, which is calculated as
\begin{equation}
    A_t^{\pi_{\theta_A^{\text{old}}}} = r(s_t, a_t) + \gamma V_{\theta_C}(s_{t+1}) - V_{\theta_C}(s_t),
\end{equation}
where $r(s_t, a_t)$ is the reward for taking action $a_t$ in state $s_t$, and $\gamma$ is the discount factor. To prevent overly large updates that may lead to unstable learning, PPO applies a clipped version of the objective function, i.e.,
\begin{equation}
\begin{aligned}
J_{\text{clip}}(\theta_A) 
&= \mathbb{E}_t \Big[ \min \Big( \rho_t(\theta_A) A_t^{\pi_{\theta_A^{\text{old}}}}, \\
&\quad \text{clip}(\rho_t(\theta_A), 1-\epsilon, 1+\epsilon) A_t^{\pi_{\theta_A^{\text{old}}}} \Big) \Big],
\end{aligned}
\end{equation}
where $\epsilon$ is a hyper-parameter that limits the update magnitude, ensuring the probability ratio $\rho_t(\theta_A)$ remains within $[1-\epsilon, 1+\epsilon]$.

\begin{remark}
To ensure stability during training and improve the sample efficiency, we incorporate generalized advantage estimation into the critic network’s value function, i.e., GAE-PPO. This technique is used to optimize the advantage function, preventing the advantage variance from growing too large during training \cite{yang2022constrained}. The generalized advantage estimation is computed as $A^{\pi_{\theta_{A}^{old}}}_t = \sum_{l=0}^{\infty} (\gamma \lambda)^l \delta_{t+l}$, where $\lambda$ is the hyperparameter introduced from GAE to adjust the number of steps of time differences considered during training. This advantage estimation ensures that the variance estimation does not diverge during training, allowing both actor and critic networks to maintain numerical stability.
\end{remark}

Next, the actor network is updated by performing gradient descent on $J_{\text{clip}}(\theta_A)$, while the critic network is updated by minimizing the mean squared error between the predicted and target value functions. The loss function is
\begin{equation}
    \mathcal{L}_C(\theta_C) = \mathbb{E}_t \left[ \left( V_{\theta_C}(s_t) - V_{\text{tar}}(s_t) \right)^2 \right],
\end{equation}
{where $V_{\theta_C}(s_t)$ is the value calculated by the value network with hyper-parameters set $\theta_C$}, and $V_{\text{tar}}(s_t)$ is the target value calculated using $n$-step returns, i.e., 
\begin{equation}
V_{\text{tar}}(s_t) = r(s_t, a_t) + \gamma V_{\theta_C}(s_{t+1}).
\end{equation}

For a clear representation of the proposed GAE-PPO method, we provide the detailed workflow in Fig.~\ref{fig:DRL}. Specifically, our GAE-PPO method provides the interaction between the actor and critic networks during the optimization process. The actor network generates actions, while the critic network evaluates the corresponding state values. Through iterative updates, the replay buffer is used to store experience samples, and the TD error is calculated to refine the value estimates. By combining advantage normalization, PPO clipping, and generalized advantage estimation, the GAE-PPO method can achieve stable and efficient training for optimizing vehicular network transmission strategies.

\subsubsection{MDP Design}

In our proposed method, the Markov decision process (MDP) framework is used to model the problem of optimizing transmission strategies for the vehicular network \cite{9354068}. The key components are designed as follows.

\textbf{Action Space:} The action space represents the decisions made by the embodied AI agents to control the transmission in the vehicular network. Specifically, the actions include channel selection, transmission power control, and the number of semantic symbols transmitted in each V2V link. The action taken by the agent at time step $t$ can be represented as
\begin{equation}
{a_t = \left( \underbrace{\{b_q[w]\}}_{Q}, \underbrace{\{P_q^{\rm V2V}[w]\}}_{Q}, \underbrace{\{u_q\}}_{Q} \right), }
\end{equation}
{Thus, the dimension of the action space is $3Q$.}

\textbf{State Space:} The state space captures the environment's observable parameters at each time step $t$. For our framework, the state $s_t$ includes the channel gain of the V2V links and V2I links, and the SINR for both V2V and V2I links. Thus, the state at time $t$ can be expressed as
\begin{equation}
{s_t = \left( \underbrace{\{\textbf{H}_i^{(w)}\}}_{W}, \underbrace{\{\gamma_q^{\text{V2V}}(t)\}}_{Q}, \underbrace{\{\gamma_w^{\text{V2I}}(t)\}}_{W} \right),}
\end{equation}
Thus, the dimension of the state space is $2W + Q$.

\textbf{Reward Function:} The reward function is designed to align with the optimization objective of the system. Specifically, our reward function encourages the maximization of the proposed QoE while considering communication constraints. The reward at time step $t$ is given by
\begin{equation}
r_t = \sum_{w=1}^W \sum_{q=1}^Q \Theta_{\rm a}\left(1 - \xi_q[w](t)\right)  \Theta_{\rm b}\left( b_q[w] P_q^{\text{V2V}}[w](t) \right).
\end{equation}
To address the challenge of ensuring that the constraints are satisfied, we employ a penalty-based reward function. Specifically, we add penalty terms to the reward function to discourage actions that violate the constraints. The designed reward function can be expressed as
\begin{equation}
\begin{aligned}
{\tilde{r}_t} &= {r_t} 
- {\lambda_1 \sum_{q=1}^Q \max\left(0, \xi_{\text{th}} - \xi_q[w](t)\right)} \\
&\quad - {\lambda_2 \sum_{q=1}^Q \max\left(0, \gamma_{\text{th}} - \gamma_q^{\text{V2V}}[w](t)\right)},
\end{aligned}
\end{equation}
where $\lambda_1$ and $\lambda_2$ are penalty coefficients that control the weight of the penalty terms. The first penalty term ensures that the semantic similarity $\xi_q[w](t)$ remains higher than or equal to the minimum threshold $\xi_{\text{th}}$, while the second penalty term ensures that the SINR $\gamma_q^{\text{V2V}}[w](t)$ remains higher than or equal to the required threshold $\gamma_{\text{th}}$. By incorporating these penalty terms, we effectively guide the embodied AI agents to make decisions that satisfy the system constraints.

\begin{algorithm}[h]
    \caption{{{GAE-PPO-Based Transmission Strategy Optimization for Vehicular Networks.}}} \label{alg:ppo_transmission}
    $\mathbf{Input:}$ Episodes, discount factor $\gamma$, PPO hyper-parameter $\epsilon$, learning rate, number of embodied AI agents;\\
    Initialize actor network $\theta_A$, critic network $\theta_C$, and environment;\\
    Observe initial state $s_0$ from environment;\\
    \For{each iteration}{
        \For{each exploration step $t$}{
            Select action $a_t$ using the actor network based on policy $\pi_{\theta_A}$;\\
            Execute action $a_t$ and observe reward $r_t$ and next state $s_{t+1}$;\\
            Store transition $(s_t, a_t, r_t, s_{t+1})$ in replay buffer $\mathcal{D}$;
        }
        \For{each update step $t$}{
            Sample mini-batch from $\mathcal{D}$;\\
            Update actor network $\theta_A$ via gradient descent on $J_{\text{clip}}(\theta_A)$;\\
            Update critic network $\theta_C$ by minimizing $\mathcal{L}_C(\theta_C)$;
        }
        Update state $s_t \gets s_{t+1}$;
    }
    $\mathbf{Output:}$ Optimized transmission policy $\pi_{\theta_A}$;
\end{algorithm}

Based on this MDP formulation, we summarize the proposed GAE-PPO-based transmission strategy optimization algorithm for vehicular networks in Algorithm 2. Specifically, the proposed algorithm utilizes the GAE to reduce variance in the policy gradient estimation, which improves the stability and efficiency of the learning process.  The algorithm begins by initializing the actor and critic networks, followed by the exploration phase, where actions are selected based on the actor's policy. The observed rewards and transitions are stored, and the actor and critic networks are updated using gradient descent to optimize the transmission policy.

\subsection{Computational Complexity Analysis}

The computational complexity of the proposed embodied AI framework is influenced by LLAVA and GAE-PPO.  For LLAVA, it primarily utilizes a Transformer-based architecture to process cross-modal data, combining visual and textual information \cite{vaswani2017attention}. The complexity of its self-attention mechanism is $\mathcal{O}(L^2 \cdot d)$, where $L$ is the input sequence length (e.g., the number of tokens or visual patches) and $d$ is the embedding dimension. Additionally, the feature projection step incurs a computational cost proportional to the projection matrix size, resulting in an overall complexity of $\mathcal{O}(L^2 \cdot d + L \cdot d^2)$.  For the GAE-PPO method, the computational complexity depends on the structure of the DNNs used in the actor and critic networks \cite{8676325}. Specifically, the complexity of the DNNs is $\mathcal{O}\left(\sum_{p=1}^P n_{p-1} \cdot n_p\right)$, where $n_p$ represents the number of neurons in the $p$-th layer, and $P$ is the total number of layers. The size of the input layer (i.e., $n_0$) corresponds to the dimension of the state space, while the size of the output layer (i.e., $n_P$) corresponds to the dimensionality of the action space.  Moreover, GAE introduces an additional computational overhead of $\mathcal{O}(T)$ per iteration, where $T$ is the number of timesteps used in advantage computation. Consequently, the overall complexity of the GAE-PPO method is as $\mathcal{O}\left(T \cdot \sum_{p=1}^P n_{p-1} \cdot n_p\right)$. Thus, the overall computational complexity of the embodied AI framework is
$\mathcal{O}(L^2 \cdot d + L \cdot d^2) + \mathcal{O}\left(T \cdot \sum_{p=1}^P n_{p-1} \cdot n_p\right)$.

\section{Numerical Results}

\subsection{Parameter Settings}
The parameters in our framework are categorized into \textit{Scenario Settings} and \textit{Embodied AI Framework Settings}.

\subsubsection{Scenario Settings}
The text dataset employed for semantic extraction in this work is the European Parliament dataset, comprising approximately 2 million sentences and 53 million words \cite{koehn2005europarl}. We simulate a vehicular network with 4 embodied AI vehicles, each supporting both V2V and V2I communication links. Following \cite{shao2024semantic}, the height of the base BS is set to 25 m, and the height of vehicles is set to 1.5 m. The V2I transmit power is fixed at 23 dBm, while the V2V transmit power varies between -100 dBm and 23 dBm. The noise powers of $\sigma_{a}^2$ and $\sigma_{b}^2$ are both set at -114 dBm, and the carrier frequency is 1 GHz. The width of road lanes is configured as 3.5 m, with shadowing deviations of 3 dB and 8 dB for V2I and V2V links, respectively. For the system objectives, {$\Theta_a (x)$ is set to $x$ and $\Theta_b$ is expressed as $\Theta_b(x)=\frac{1}{1+e^{-x}}$}. The minimum thresholds for SSIM and SINR constraints are set to $\xi_{\rm th}=0.3$ and $\gamma_{\rm th}=10 \rm~ dB$, respectively.

\subsubsection{Embodied AI Framework Settings}
The framework consists of two key components, i.e.,  LLAVA and GAE-PPO. Specifically, the LLAVA model used in our simulations is LLAVA-v1.5-7B\footnote{\url{https://huggingface.co/llava-hf/llava-1.5-7b-hf}}, containing 7 billion adjustable parameters. It processes a dataset comprising 30 driving scenario images \cite{liu2024improved}, efficiently extracting semantic information to support vehicular communication. {The transmitter network includes a semantic encoder and a channel encoder, with hyper-parameters $\alpha$ and $\beta$. The architecture consists of three Transformer layers, each with 8 attention heads, and the Rectified Linear Unit (ReLU) as the activation function. Two dense layers, with sizes of 256 and 16, are connected to the last Transformer layer. The receiver network, comprising a semantic decoder and a channel decoder, is similarly structured but with the layers applied in reverse order.}  For the GAE-PPO algorithm, we use the \textit{Adam} optimizer and \textit{Tanh} as the activation function. The neural network includes three layers, each with 512 hidden units \cite{10702596}. {The replay buffer has a maximum size of $1 \times 10^{6}$, and the network is updated at the end of each episode.} The penalty coefficients $\lambda_1$ and $\lambda_2$ are both set to 1. Other critical parameters are listed in Table~\ref{tab_PS} \cite{10181138}.

\begin{table}[!t]
\begin{small}
\begin{center}
\caption{Critical Parameter Settings}
\label{tab_PS}
\centering
\begin{tabular}{|c|c|}  
\hline
\textbf{System Parameter}  & \textbf{Value} \\ \hline \hline
V2I transmit power & 23 dBm \\ \hline
V2V transmit power & [-100, 5, 15, 23] dBm \\ \hline
V2I deviation of shadowing & 3 dB \\ \hline
V2V deviation of shadowing & 8 dB \\ \hline
Number of NN layers & 3 \\ \hline
Clipping parameter $\epsilon$ & 0.5 \\ \hline
Initial learning rate $\alpha$ & $1 \times 10^{-4}$ \\ \hline
Final learning rate $\alpha$ & $1 \times 10^{-8}$ \\ \hline
Discount factor $\gamma$ & 0.99 \\ \hline

\end{tabular}
\end{center}
\end{small}
\end{table}



\begin{figure*}[!t]
\centering
\includegraphics[width=0.95\textwidth]{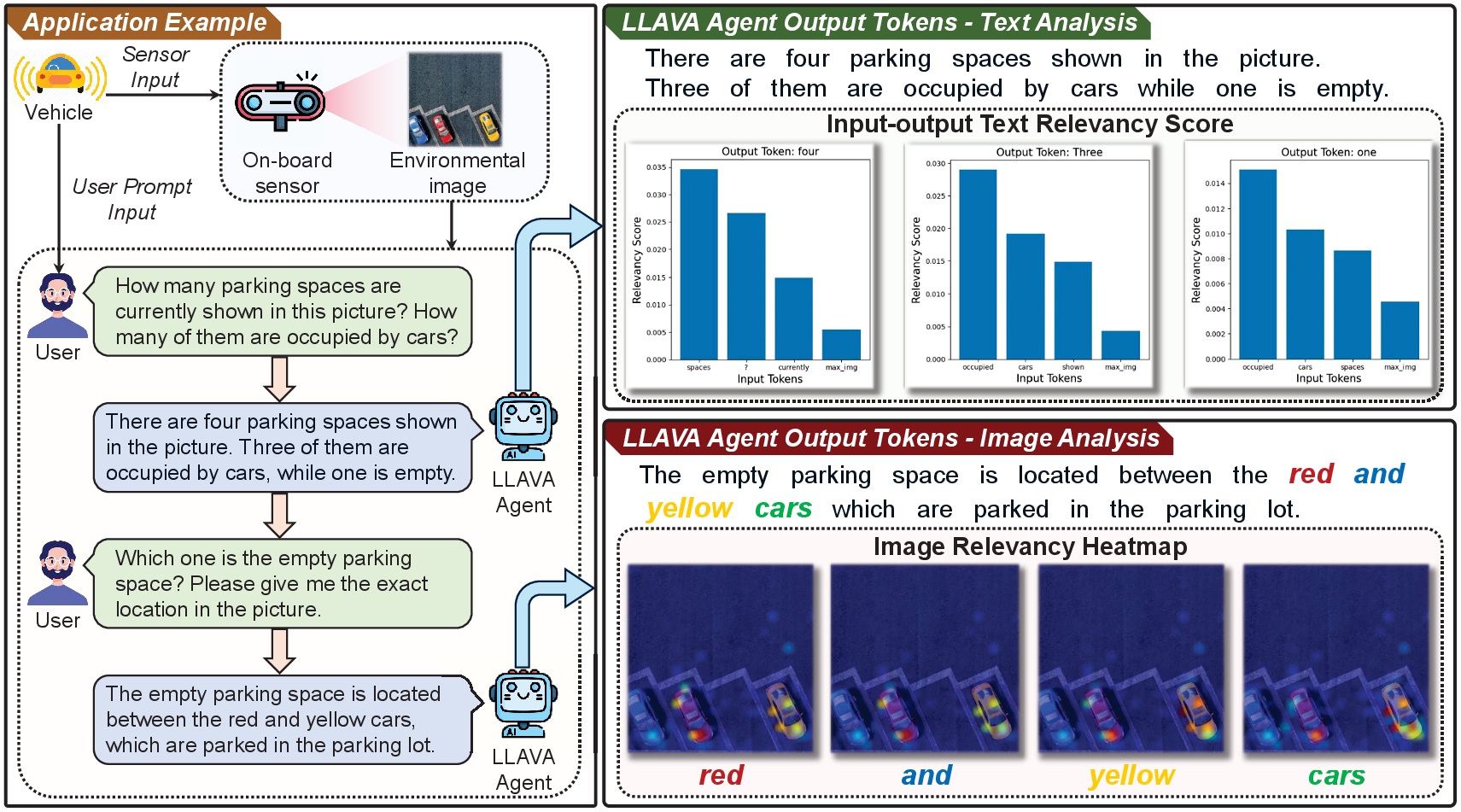}
\caption{The illustration of semantic information extraction.
In the left part, the environmental image is obtained through the onboard sensor. Then, the semantic information is extracted according to the user's needs via the embodied AI. In the right part, we demonstrate the relevance between the extracted semantic information and user needs, as well as the environmental images.
}
\label{fig:enter-label2}
\end{figure*}

\subsection{Simulation Results}
Fig.~\ref{fig:enter-label2} demonstrates an example of semantic information extraction when an embodied AI vehicle captures images of its surroundings. As shown, thanks to LLAVA's powerful cross-modal capabilities, the system can effectively extract key details such as the number of parking spaces, which ones are occupied, and where available parking spots are located. Compared to transmitting the image directly (i.e., 614 Kilobyte in our simulations), sending the extracted text (i.e., 12.1 Kilobyte in our simulations) significantly reduces the required bandwidth. This highlights the efficiency of using text-based semantic information over raw image data for communication purposes in vehicular networks. Moreover, the right part of Fig.~\ref{fig:enter-label2} presents a detailed analysis of the semantic extraction process, focusing on LLAVA's attention mechanism and how the textual output is associated with visual elements. This analysis demonstrates how LLAVA combines input text with visual features to generate the final output, ensuring that the extracted information is well-grounded in the environmental context.
\begin{itemize} \item \textbf{Input-Output Text Relevancy Analysis:} The bar charts show the relevancy scores between input tokens and output tokens, reflecting the contribution of each input word to generating specific output tokens. Tokens such as "spaces," "occupied," and "cars" are found to be highly relevant to responses like "four," "three," and "one." This indicates that LLAVA can effectively direct its attention to specific, relevant parts of the input to generate accurate output, ensuring the correct interpretation of semantic elements, such as the number of parking spaces and their occupancy status.

\item \textbf{Relevancy Heatmaps:} The heatmaps at the bottom of Fig.~\ref{fig:enter-label2}  highlight the image regions most relevant to specific output tokens. For instance, the highlighted areas show that tokens "red" and "yellow" are directly associated with the respective cars in the image. This demonstrates LLAVA's capability to ground text in specific visual features, linking language tokens to corresponding visual objects. The heatmaps validate the system's ability to perform cross-modal grounding, which enhances the reliability of extracted semantic information by correctly associating textual descriptions with visual inputs.
\end{itemize}

\begin{figure*}[!t]
\centering
\includegraphics[width=0.95\textwidth]{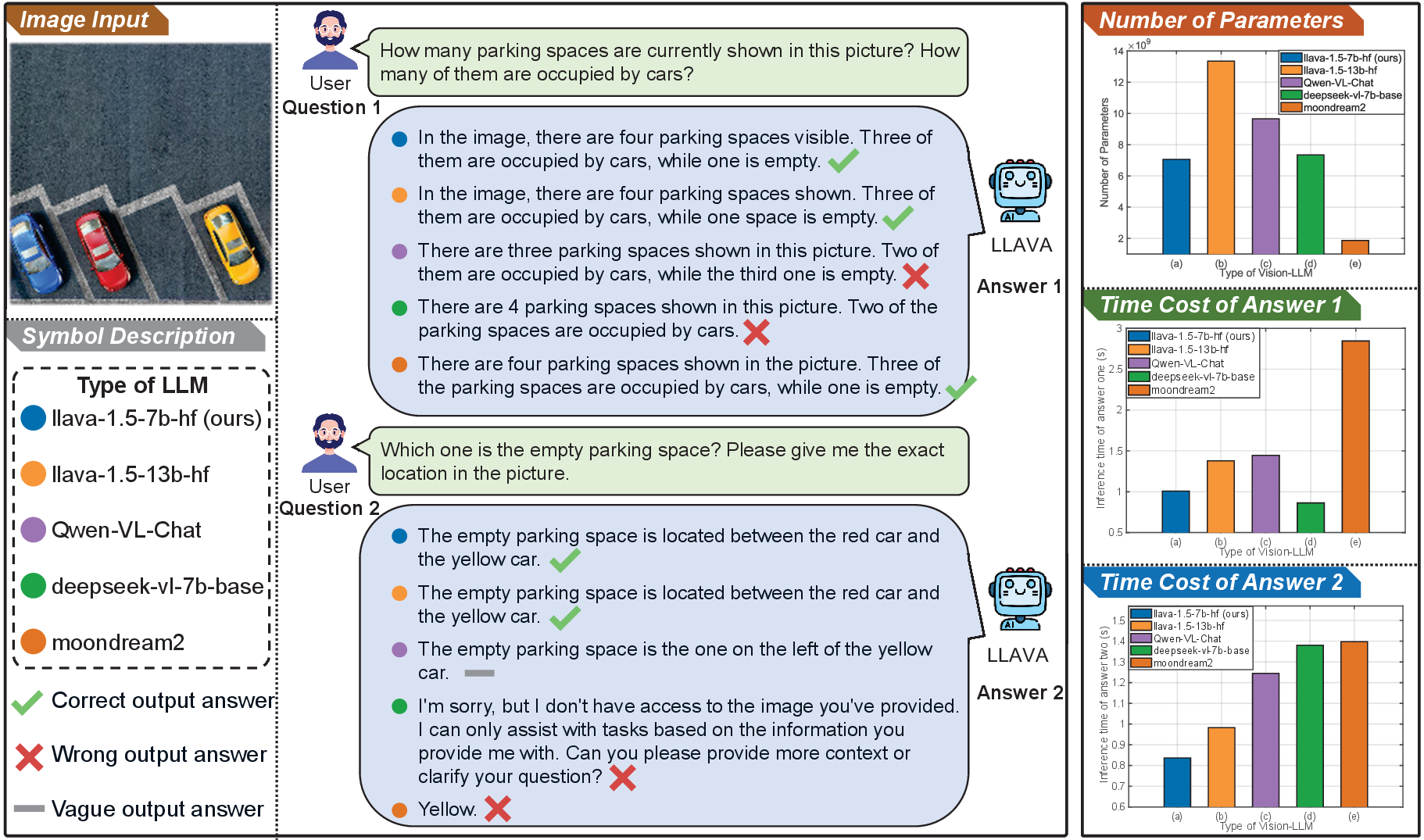}
\caption{Comparison of vision-LLM models in terms of parameter size, inference time, and output accuracy for semantic queries. The left side illustrates the semantic query results for two questions related to parking spaces, highlighting the correctness of answers provided by different models. The bar charts on the right compare the parameter sizes and inference times for each model, showing the balance between computational efficiency and output accuracy. All experiments are conducted on a server equipped with an Intel Xeon Platinum 8380 CPU and NVIDIA A100 80GB GPU.}
\label{fig:enter-label21}
\end{figure*}

Next, to evaluate the performance of our LLAVA-1.5-7b-hf model, we compare it with several state-of-the-art vision-LLM models, including:
\begin{itemize}
    \item \textbf{LLAVA-1.5-13b-hf} \cite{lin2023video}: A vision-LLM with about 13.4 billion parameters, designed for high-accuracy multimodal semantic extraction through advanced cross-modal alignment techniques.
    
    \item \textbf{Qwen-VL-Chat} \cite{bai2023qwen}: A multimodal conversational model with about 9.66 billion parameters, specializing in visual-language reasoning tasks requiring both image and text understanding.
    
    \item \textbf{Deepseek-vl-7b-base} \cite{lu2024deepseek}: A vision-LLM with about 7.34 billion parameters optimized for extracting structured semantic information from complex visual and textual data.
    
    \item \textbf{Moondream2} \cite{murthy2024mobileaibench}: A lightweight vision-LLM with 1.8 billion parameters, targeting general-purpose semantic reasoning with the focus on computational efficiency.
\end{itemize}
Fig.~\ref{fig:enter-label21} compares these vision-LLM models in terms of parameter size, inference time, and output accuracy for semantic queries, highlighting their trade-offs in computational efficiency and performance. The first bar chart shows the parameter sizes of the evaluated models. LLAVA-1.5-7b-hf, with 7 billion parameters, is significantly smaller than LLAVA-1.5-13b-hf, which has 13 billion parameters, approximately 46.2\% fewer parameters. Despite this reduction in size, LLAVA-1.5-7b-hf achieves comparable or superior performance, making it particularly suitable for real-time decision-making in resource-constrained vehicular networks. The second and third bar charts evaluate inference time for two semantic queries, both requiring the identification of parking space availability and its exact location. LLAVA-1.5-7b-hf achieves inference times of 1.2 seconds and 1.3 seconds for the first and second queries, respectively, which are 40\% and 38\% faster compared to LLAVA-1.5-13b-hf. While LLAVA-1.5-13b-hf delivers accurate answers, its larger parameter size results in higher latency. Other models, such as Qwen-VL-Chat and deepseek-vl-7b-base, either fail to provide consistent, correct answers or exhibit even higher inference delays, further emphasizing the efficiency of LLAVA-1.5-7b-hf. On the left side of Fig.~\ref{fig:enter-label21}, LLAVA-1.5-7b-hf demonstrates robust semantic alignment between visual inputs and textual representations. For instance, it consistently produces correct answers for both semantic queries, while the other models either provide vague or incorrect responses. This performance advantage can be attributed to its efficient cross-modal alignment mechanism, which balances parameter efficiency and semantic understanding. These characteristics make it a strong candidate for embodied AI vehicular networks, where low-latency and high-reliability decision-making are essential.

\begin{figure}[!t]
\centering
\includegraphics[width=0.49\textwidth]{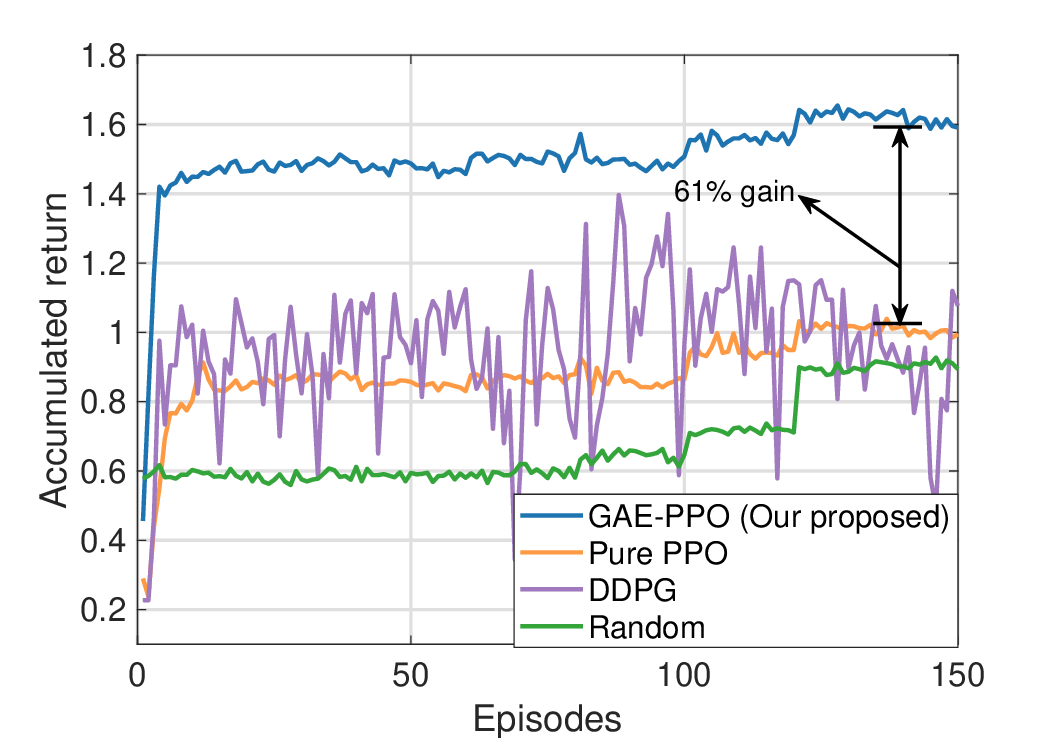}
\caption{Convergence behavior with different methods.}
\label{fig:enter-label3}
\end{figure}

\begin{figure}[!t]
\centering
\includegraphics[width=0.49\textwidth]{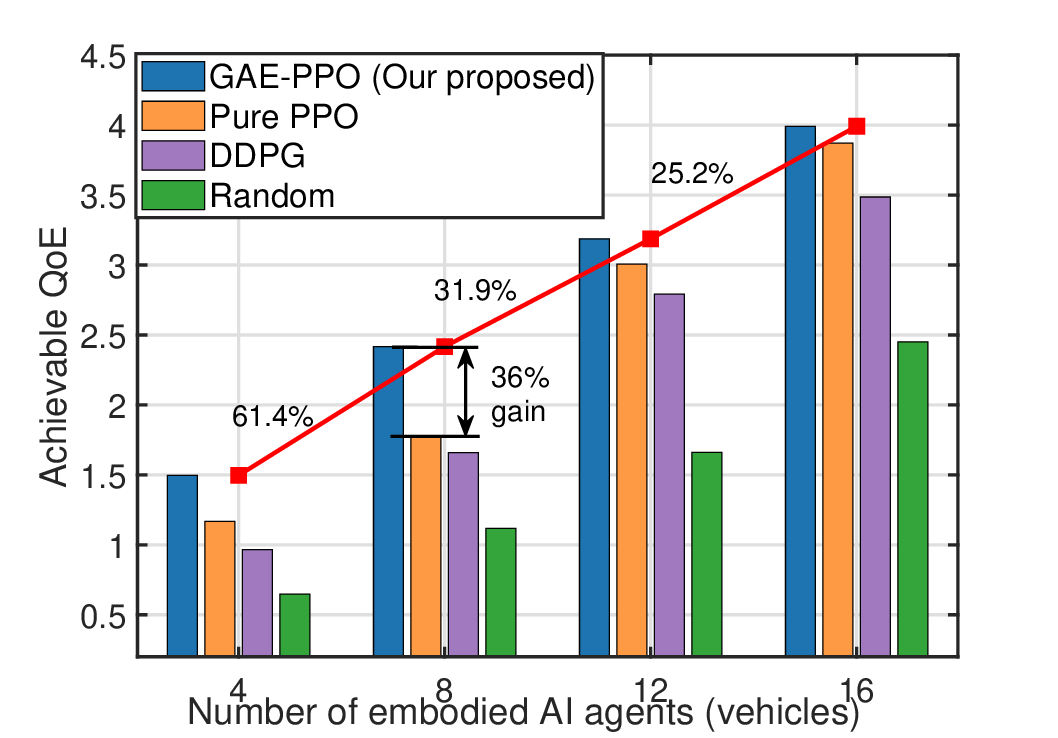}
\caption{Achievable QoE versus the number of embodied AI agents.}
\label{fig:enter-label4}
\end{figure}


\begin{figure*}[htbp]
\centering
\begin{subfigure}{.24\textwidth}
  \centering
  \includegraphics[width=1.0\linewidth]{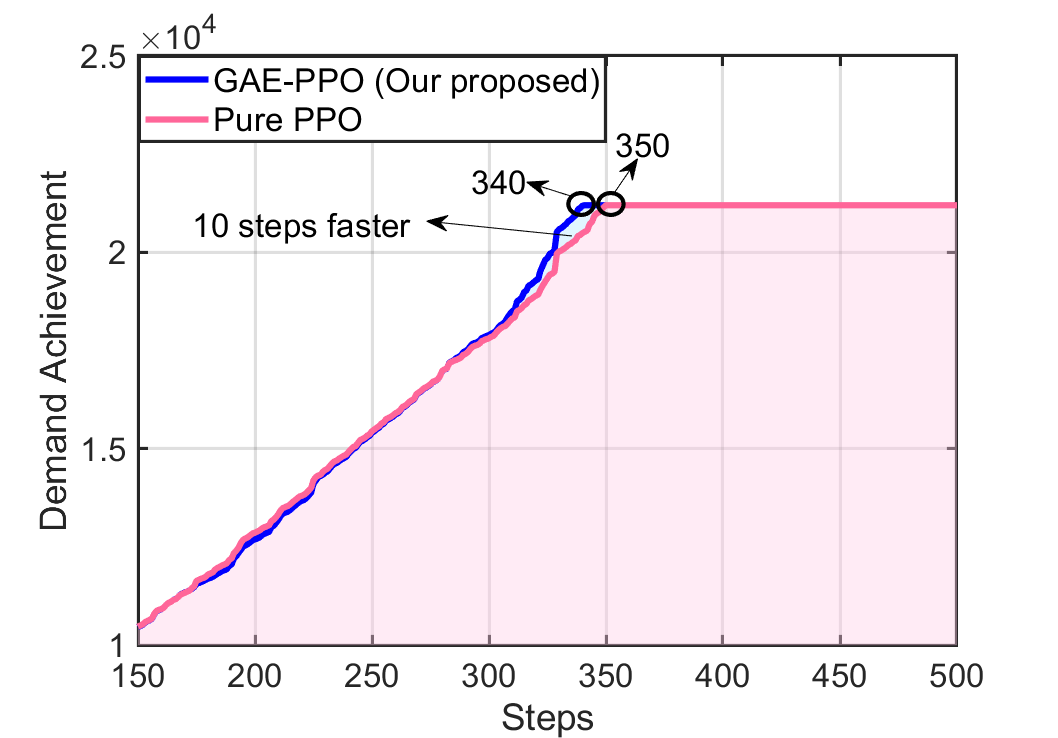} 
  \caption{1\textsuperscript{st} vehicle's demand}
  \label{fig:sub1}
\end{subfigure}%
\begin{subfigure}{.24\textwidth}
  \centering
  \includegraphics[width=1.0\linewidth]{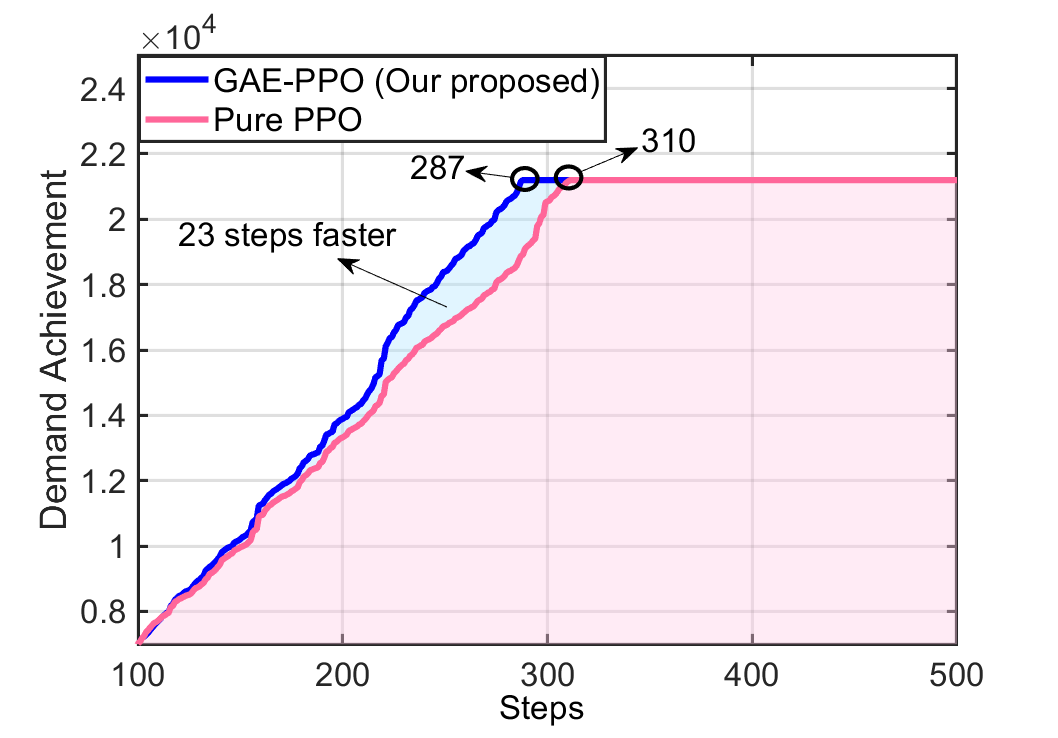}
  \caption{2\textsuperscript{nd} vehicle's demand}
  \label{fig:sub2}
\end{subfigure}
\begin{subfigure}{.24\textwidth}
  \centering
  \includegraphics[width=1.0\linewidth]{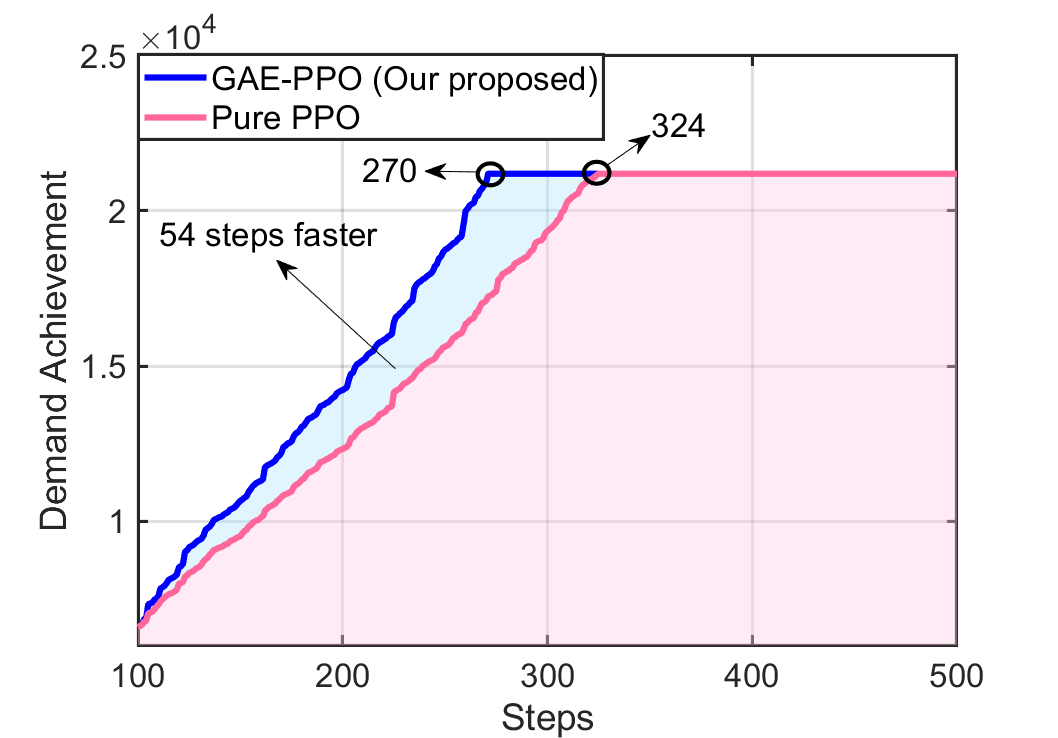}
  \caption{3\textsuperscript{rd} vehicle's demand}
  \label{fig:sub3}
\end{subfigure}
\begin{subfigure}{.24\textwidth}
  \centering
  \includegraphics[width=1.0\linewidth]{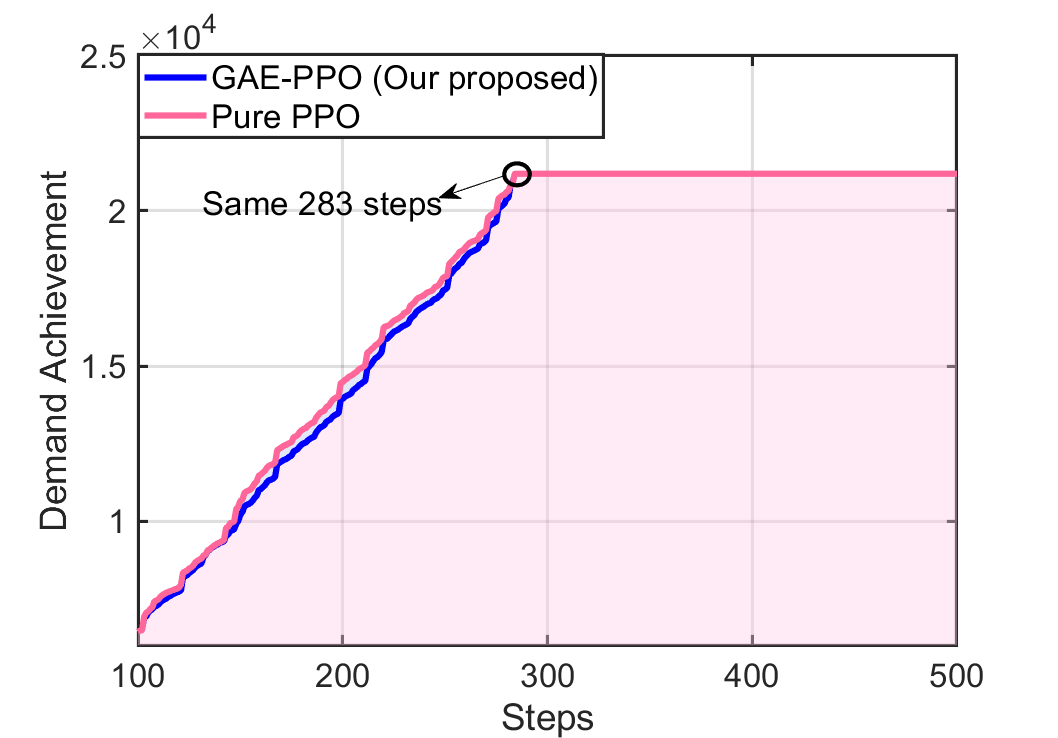}
  \caption{4\textsuperscript{th} vehicle's demand}
  \label{fig:sub4}
\end{subfigure}%
\caption{Comparison of GAE-PPO and pure PPO in fulfilling communication demands.}
\label{fig:comparison}
\end{figure*}

\begin{figure}[!t]
\centering
\includegraphics[width=0.49\textwidth]{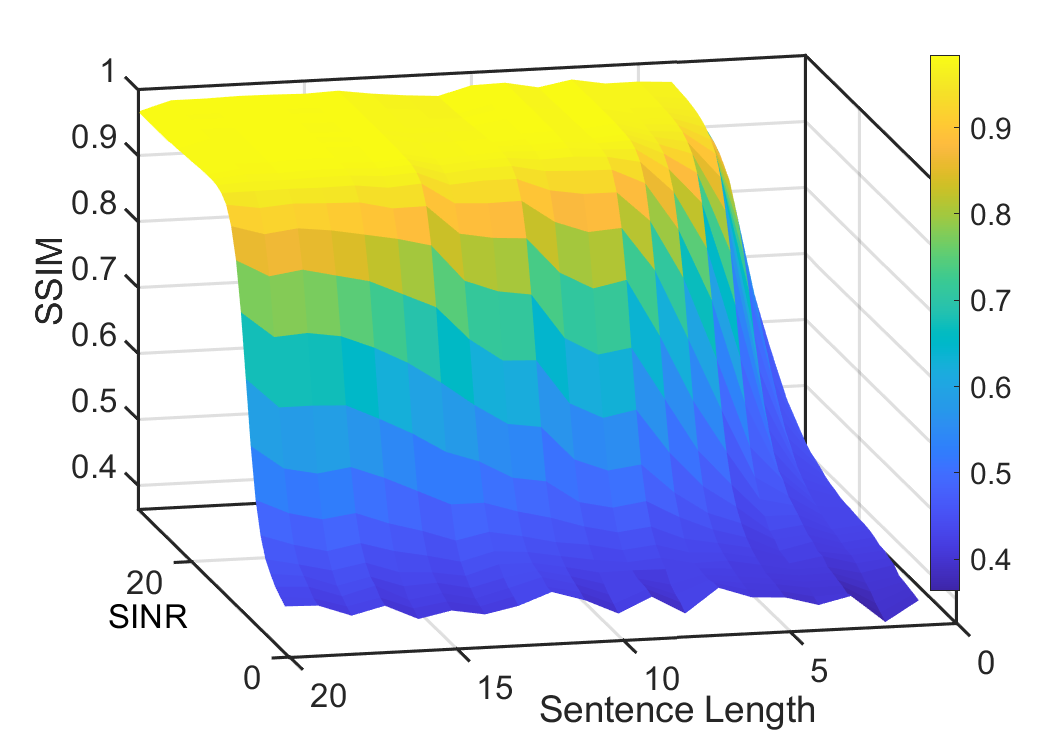}
\caption{The SSIM versus different words sentence length and SINR.}
\label{fig:radar}
\end{figure}

After extracting semantic information, the proposed GAE-PPO method is used to transmit the data. Fig.~\ref{fig:enter-label3} shows a convergence comparison of our GAE-PPO method against several baselines, i.e., pure PPO, DDPG, and a random policy. The results indicate that our GAE-PPO method achieves superior performance compared to pure PPO, providing approximately 61\% gain in accumulated return. This improvement is attributed to the GAE mechanism, which stabilizes the learning process and results in more effective policy updates. Compared to DDPG, which demonstrates significant variability with fluctuations, the method struggles with consistent convergence, indicating its instability and unsuitability for reliable performance. In contrast, our GAE-PPO method leverages generalized advantage estimation, which significantly contributes to stable convergence behavior, making it more reliable in vehicular network environments.

Fig.~\ref{fig:enter-label4} shows the achievable QoE performance of different transmission strategies as the number of embodied AI agents (i.e., vehicles) varies. The proposed GAE-PPO method consistently achieves superior performance compared to DDPG, pure PPO, and a random strategy across all scenarios. Specifically, our GAE-PPO method demonstrates a 36\% gain in QoE over DDPG at a vehicle count of 8, highlighting its effectiveness in dynamic vehicular environments. The red line in Fig.~\ref{fig:enter-label4} illustrates the QoE improvements as the number of vehicles increases. Notably, we observe a 61.4\% increase in QoE when scaling from 4 to 8 vehicles, followed by a 31.9\% improvement from 8 to 12 vehicles and a 25.2\% increase when moving from 12 to 16 vehicles. These results demonstrate the scalability of our GAE-PPO method, as it maintains consistent performance gains even as the network size grows. The observed improvements can be largely attributed to the use of GAE, which provides more accurate advantage estimation, leading to smoother learning and better resource utilization.

To understand the advantages of our proposed method, Fig.~\ref{fig:comparison} illustrates the successfully delivered payloads for each vehicle of GAE-PPO and pure PPO methods across four embodied AI agents (i.e., vehicles), highlighting the number of steps each method requires to meet communication demands. It shows that GAE-PPO consistently outperforms pure PPO by achieving faster convergence for most vehicles. Specifically, GAE-PPO reduces the required steps by 10, 23, and 54 for vehicles 1, 2, and 3, respectively. For vehicle 4, both methods achieve the communication demand in the same number of steps, indicating comparable performance in this scenario.  The efficiency of GAE-PPO is attributed to its generalized advantage estimation mechanism, which stabilizes the training process and enhances the balance between exploration and exploitation. This enables GAE-PPO to adapt effectively to dynamic vehicular communication environments, prioritizing optimal actions that lead to faster demand fulfillment.

To explore the impact of dataset size, Fig.~\ref{fig:radar} illustrates the relationship between the SSIM, the SINR, and the semantic sentence length. It shows that SSIM is significantly influenced by both SINR and sentence length, reflecting how these factors affect the quality of semantic communication in the vehicular network. At higher SINR values, the SSIM remains consistently high regardless of the sentence length, indicating that a reliable channel effectively preserves the quality of transmitted semantic information, even with longer sentences. However, as SINR decreases, we observe a notable drop in SSIM, particularly for longer sentences. This suggests that in low-SINR environments, transmitting longer semantic information leads to a greater degradation in quality due to increased noise interference. Moreover, it also indicates that shorter sentence lengths tend to maintain higher SSIM values across varying SINR conditions. This highlights the importance of optimizing sentence length to ensure reliable communication quality, especially in challenging network conditions where SINR is low. These observations underscore the trade-off between semantic content richness and communication quality and suggest that dynamically adapting sentence length based on SINR can be an effective strategy to enhance the performance of embodied AI-enhanced vehicular networks.

\section{Conclusions}
In this paper, we have proposed a framework for optimizing transmission and decision-making in embodied AI-enhanced vehicular networks. By integrating LLAVA for semantic information extraction and GAE-PPO for stable decision-making, the proposed method can improve transmission efficiency and adaptability in dynamic vehicular environments. We have also utilized attention maps from LLAVA to analyze how semantic information is extracted, demonstrating LLAVA's ability to focus on the most relevant visual regions for accurate semantic representation. The results have demonstrated that our method consistently outperforms traditional methods in terms of convergence speed, transmission efficiency, and overall system performance, validating the effectiveness of using embodied AI framework in vehicular networks.

\bibliographystyle{IEEEtran}
\bibliography{IEEEabrv, Bibliography}

\begin{thebibliography}{10}
\providecommand{\url}[1]{#1}
\csname url@rmstyle\endcsname
\providecommand{\newblock}{\relax}
\providecommand{\bibinfo}[2]{#2}
\providecommand\BIBentrySTDinterwordspacing{\spaceskip=0pt\relax}
\providecommand\BIBentryALTinterwordstretchfactor{4}
\providecommand\BIBentryALTinterwordspacing{\spaceskip=\fontdimen2\font plus
\BIBentryALTinterwordstretchfactor\fontdimen3\font minus \fontdimen4\font\relax}
\providecommand\BIBforeignlanguage[2]{{%
\expandafter\ifx\csname l@#1\endcsname\relax
\typeout{** WARNING: IEEEtran.bst: No hyphenation pattern has been}%
\typeout{** loaded for the language `#1'. Using the pattern for}%
\typeout{** the default language instead.}%
\else
\language=\csname l@#1\endcsname
\fi
#2}}

\bibitem{9815151}
S.~B. Prathiba, G.~Raja, S.~Anbalagan, \emph{et~al.}, ``A hybrid deep sensor anomaly detection for autonomous vehicles in 6{G}-{V2X} environment,'' \emph{IEEE Trans. Network Sci. Eng.}, vol.~10, no.~3, pp. 1246--1255, 2023.

\bibitem{10571385}
A.~Nahar, K.~K. Mondal, D.~Das, \emph{et~al.}, ``Clouds on the road: A software-defined fog computing framework for intelligent resource management in vehicular {A}d-{H}oc networks,'' \emph{IEEE Trans. Mob. Comput.}, vol.~23, no.~12, pp. 12\,778--12\,792, 2024.

\bibitem{wang2024privacy}
L.~Wang, H.~Zhong, J.~Cui, \emph{et~al.}, ``Privacy-preserving and secure distributed data sharing scheme for {VANET}s,'' \emph{IEEE Trans. Mob. Comput.}, 2024.

\bibitem{liu2024fedagl}
S.~Liu, Y.~Li, P.~Guan, \emph{et~al.}, ``Fedagl: A communication-efficient federated vehicular network,'' \emph{IEEE Trans. Intell. Veh.}, 2024.

\bibitem{wang2023reliable}
Z.~Wang, Y.~Liu, J.~Wang, \emph{et~al.}, ``A reliable physical layer key generation scheme based on {RSS} and {LSTM} network in {VANET},'' \emph{IEEE Internet Things J.l}, vol.~11, no.~1, pp. 692--707, 2023.

\bibitem{tang2021comprehensive}
F.~Tang, B.~Mao, N.~Kato, \emph{et~al.}, ``Comprehensive survey on machine learning in vehicular network: Technology, applications and challenges,'' \emph{IEEE Commun. Survey Tuts.}, vol.~23, no.~3, pp. 2027--2057, 2021.

\bibitem{duan2022survey}
J.~Duan, S.~Yu, H.~L. Tan, \emph{et~al.}, ``A survey of embodied {AI}: From simulators to research tasks,'' \emph{IEEE Trans. Emerging Top. Comput. Intell.}, vol.~6, no.~2, pp. 230--244, 2022.

\bibitem{cunneen2019autonomous}
M.~Cunneen, M.~Mullins, and F.~Murphy, ``Autonomous vehicles and embedded artificial intelligence: The challenges of framing machine driving decisions,'' \emph{Applied Artificial Intelligence}, vol.~33, no.~8, pp. 706--731, 2019.

\bibitem{10679152}
R.~Zhang, H.~Du, Y.~Liu, \emph{et~al.}, ``Generative {AI} agents with large language model for satellite networks via a mixture of experts transmission,'' \emph{IEEE J. Sel. Areas Commun.}, pp. 1--1, 2024.

\bibitem{wang2024omnidrive}
S.~Wang, Z.~Yu, X.~Jiang, \emph{et~al.}, ``Omnidrive: A holistic {LLM}-agent framework for autonomous driving with 3d perception, reasoning and planning,'' \emph{arXiv preprint arXiv:2405.01533}, 2024.

\bibitem{10032267}
R.~Zhang, K.~Xiong, Y.~Lu, \emph{et~al.}, ``Energy efficiency maximization in {RIS}-assisted {SWIPT} networks with {RSMA}: A {PPO}-based approach,'' \emph{IEEE J. Sel. Areas Commun.}, vol.~41, no.~5, pp. 1413--1430, 2023.

\bibitem{10506539}
R.~Zhang, K.~Xiong, H.~Du, \emph{et~al.}, ``Generative {AI}-enabled vehicular networks: Fundamentals, framework, and case study,'' \emph{IEEE Netw.}, vol.~38, no.~4, pp. 259--267, 2024.

\bibitem{shao2024semantic}
Z.~Shao, Q.~Wu, P.~Fan, \emph{et~al.}, ``Semantic-aware spectrum sharing in internet of vehicles based on deep reinforcement learning,'' \emph{arXiv preprint arXiv:2406.07213}, 2024.

\bibitem{10531073}
R.~Zhang, H.~Du, Y.~Liu, \emph{et~al.}, ``Interactive ai with retrieval-augmented generation for next generation networking,'' \emph{IEEE Netw.}, vol.~38, no.~6, pp. 414--424, 2024.

\bibitem{liang2019spectrum}
L.~Liang, H.~Ye, and G.~Y. Li, ``Spectrum sharing in vehicular networks based on multi-agent reinforcement learning,'' \emph{IEEE J. Sel. Areas Commun.}, vol.~37, no.~10, pp. 2282--2292, 2019.

\bibitem{liang2018graph}
L.~Liang, S.~Xie, G.~Y. Li, \emph{et~al.}, ``Graph-based resource sharing in vehicular communication,'' \emph{IEEE Trans. Wireless Commun.}, vol.~17, no.~7, pp. 4579--4592, 2018.

\bibitem{xue2024cooperative}
J.~Xue, K.~Yu, T.~Zhang, \emph{et~al.}, ``Cooperative deep reinforcement learning enabled power allocation for packet duplication {URLLC} in multi-connectivity vehicular networks,'' \emph{IEEE Trans. Mob. Comput.}, 2024.

\bibitem{10643168}
C.~Zhang, W.~Zhang, Q.~Wu, \emph{et~al.}, ``Distributed deep reinforcement learning based gradient quantization for federated learning enabled vehicle edge computing,'' \emph{IEEE Internet Things J.l}, pp. 1--1, 2024.

\bibitem{yao2023secure}
Y.~Yao, F.~Shu, Z.~Li, \emph{et~al.}, ``Secure transmission scheme based on joint radar and communication in mobile vehicular networks,'' \emph{IEEE Trans. Intell. Transp. Syst.}, vol.~24, no.~9, pp. 10\,027--10\,037, 2023.

\bibitem{savva2019habitat}
M.~Savva, A.~Kadian, O.~Maksymets, \emph{et~al.}, ``Habitat: A platform for embodied ai research,'' in \emph{Proc. IEEE ICCV}, 2019, pp. 9339--9347.

\bibitem{song2023llm}
C.~H. Song, J.~Wu, C.~Washington, \emph{et~al.}, ``{LLM}-planner: Few-shot grounded planning for embodied agents with large language models,'' in \emph{Proc. IEEE ICCV}, 2023, pp. 2998--3009.

\bibitem{zhang2024towards}
Y.~Zhang, S.~Yang, C.~Bai, \emph{et~al.}, ``Towards efficient {LLM} grounding for embodied multi-agent collaboration,'' \emph{arXiv preprint arXiv:2405.14314}, 2024.

\bibitem{mower2024ros}
C.~E. Mower, Y.~Wan, H.~Yu, \emph{et~al.}, ``Ros-{LLM}: A {ROS} framework for embodied {AI} with task feedback and structured reasoning,'' \emph{arXiv preprint arXiv:2406.19741}, 2024.

\bibitem{zhang2024badrobot}
H.~Zhang, C.~Zhu, X.~Wang, \emph{et~al.}, ``Bad{R}obot: Jailbreaking llm-based embodied {AI} in the physical world,'' \emph{arXiv preprint arXiv:2407.20242}, 2024.

\bibitem{Shunpu_SemCom}
S.~Tang, Q.~Yang, L.~Fan, \emph{et~al.}, ``Contrastive learning-based semantic communications,'' \emph{IEEE Trans. Commun.}, vol.~72, no.~10, pp. 6328--6343, 2024.

\bibitem{ying2024peac}
C.~Ying, Z.~Hao, X.~Zhou, \emph{et~al.}, ``{PEAC}: Unsupervised pre-training for cross-embodiment reinforcement learning,'' \emph{arXiv preprint arXiv:2405.14073}, 2024.

\bibitem{10146515}
Y.~Long, W.~Wei, T.~Huang, \emph{et~al.}, ``Human-in-the-loop embodied intelligence with interactive simulation environment for surgical robot learning,'' \emph{IEEE Robot. Autom. Lett.}, vol.~8, no.~8, pp. 4441--4448, 2023.

\bibitem{liu2023simple}
E.~Z. Liu, S.~Suri, T.~Mu, \emph{et~al.}, ``Simple embodied language learning as a byproduct of meta-reinforcement learning,'' in \emph{Proc. ICML}.\hskip 1em plus 0.5em minus 0.4em\relax PMLR, 2023, pp. 21\,997--22\,008.

\bibitem{tan2024true}
W.~Tan, W.~Zhang, S.~Liu, \emph{et~al.}, ``True knowledge comes from practice: Aligning llms with embodied environments via reinforcement learning,'' \emph{arXiv preprint arXiv:2401.14151}, 2024.

\bibitem{10041763}
B.~Shin, J.~H. Lee, C.~Kim, \emph{et~al.}, ``Lte rssi based vehicular localization system in long tunnel environment,'' \emph{IEEE Trans. Ind. Informat.}, vol.~19, no.~11, pp. 11\,102--11\,114, 2023.

\bibitem{huang2024joint}
M.~Huang, Z.~Shen, and G.~Zhang, ``Joint spectrum sharing and {V2V}/{V2I} task offloading for vehicular edge computing networks based on coalition formation game,'' \emph{IEEE Trans. Intell. Transp. Syst.}, 2024.

\bibitem{9808399}
F.~Liu, J.~Chen, Q.~Zhang, \emph{et~al.}, ``Online {MEC} offloading for {V2V} networks,'' \emph{IEEE Trans. Mob. Comput.}, vol.~22, no.~10, pp. 6097--6109, 2023.

\bibitem{xie2021deep}
H.~Xie, Z.~Qin, G.~Y. Li, \emph{et~al.}, ``Deep learning enabled semantic communication systems,'' \emph{IEEE Trans. Signal Process.}, vol.~69, pp. 2663--2675, 2021.

\bibitem{10753492}
K.~Li, J.~Zheng, W.~Ni, \emph{et~al.}, ``Biasing federated learning with a new adversarial graph attention network,'' \emph{IEEE Trans. Mob. Comput.}, pp. 1--15, 2024.

\bibitem{10746594}
R.~Cheng, Y.~Sun, D.~Niyato, \emph{et~al.}, ``A wireless {AI}-generated content ({AIGC}) provisioning framework empowered by semantic communication,'' \emph{IEEE Trans. Mob. Comput.}, pp. 1--14, 2024.

\bibitem{10654734}
D.~Sheng, Q.~Qi, J.~Wang, \emph{et~al.}, ``Psyqoe: Improving quality-of-experience assessment with psychological effects in video streaming,'' \emph{IEEE Trans. Serv. Comput.}, pp. 1--14, 2024.

\bibitem{xie2023ra}
C.-W. Xie, S.~Sun, X.~Xiong, \emph{et~al.}, ``Ra-clip: Retrieval augmented contrastive language-image pre-training,'' in \emph{Proc. IEEE CVPR}, 2023, pp. 19\,265--19\,274.

\bibitem{liu2024visual}
H.~Liu, C.~Li, Q.~Wu, \emph{et~al.}, ``Visual instruction tuning,'' \emph{NIPS}, vol.~36, 2024.

\bibitem{ji2022trajectory}
J.~Ji, K.~Zhu, and L.~Cai, ``Trajectory and communication design for cache-enabled {UAV}s in cellular networks: A deep reinforcement learning approach,'' \emph{IEEE Trans. Mob. Comput.}, vol.~22, no.~10, pp. 6190--6204, 2022.

\bibitem{yang2022constrained}
L.~Yang, J.~Ji, J.~Dai, \emph{et~al.}, ``Constrained update projection approach to safe policy optimization,'' \emph{NIPS}, vol.~35, pp. 9111--9124, 2022.

\bibitem{9354068}
C.~Li, J.~Xia, F.~Liu, \emph{et~al.}, ``Dynamic offloading for multiuser muti-{CAP} mec networks: A deep reinforcement learning approach,'' \emph{IEEE Trans. Veh. Technol.}, vol.~70, no.~3, pp. 2922--2927, 2021.

\bibitem{vaswani2017attention}
A.~Vaswani, ``Attention is all you need,'' \emph{NIPS}, 2017.

\bibitem{8676325}
C.~H. Liu, X.~Ma, X.~Gao, \emph{et~al.}, ``Distributed energy-efficient multi-{UAV} navigation for long-term communication coverage by deep reinforcement learning,'' \emph{IEEE Trans. Mob. Comput.}, vol.~19, no.~6, pp. 1274--1285, 2020.

\bibitem{koehn2005europarl}
P.~Koehn, ``Europarl: A parallel corpus for statistical machine translation,'' in \emph{Proceedings of machine translation summit x: papers}, 2005, pp. 79--86.

\bibitem{liu2024improved}
H.~Liu, C.~Li, Y.~Li, \emph{et~al.}, ``Improved baselines with visual instruction tuning,'' in \emph{Proc. IEEE CVPR}, 2024, pp. 26\,296--26\,306.

\bibitem{10702596}
X.~He, Y.~Yang, J.~Lee, \emph{et~al.}, ``Deep reinforcement learning based aoi minimization for {NOMA}-enabled integrated satellite-terrestrial networks,'' \emph{IEEE Trans. Veh. Technol.}, pp. 1--6, 2024.

\bibitem{10181138}
T.~Zhang, K.-Y. Lam, J.~Zhao, \emph{et~al.}, ``Joint device scheduling and bandwidth allocation for federated learning over wireless networks,'' \emph{IEEE Trans. Wireless Commun.}, pp. 1--1, 2023.

\bibitem{lin2023video}
B.~Lin, Y.~Ye, B.~Zhu, \emph{et~al.}, ``Video-llava: Learning united visual representation by alignment before projection,'' \emph{arXiv preprint arXiv:2311.10122}, 2023.

\bibitem{bai2023qwen}
J.~Bai, S.~Bai, S.~Yang, \emph{et~al.}, ``{Q}wen-vl: A frontier large vision-language model with versatile abilities,'' \emph{arXiv preprint arXiv:2308.12966}, 2023.

\bibitem{lu2024deepseek}
H.~Lu, W.~Liu, B.~Zhang, \emph{et~al.}, ``Deepseek-vl: towards real-world vision-language understanding,'' \emph{arXiv preprint arXiv:2403.05525}, 2024.

\bibitem{murthy2024mobileaibench}
R.~Murthy, L.~Yang, J.~Tan, \emph{et~al.}, ``Mobileaibench: Benchmarking {LLM}s and {LMM}s for on-device use cases,'' \emph{arXiv preprint arXiv:2406.10290}, 2024.

\end{thebibliography}

\end{document}